%% file: main.tex
\documentclass[useAMS,usenatbib,twocolumn,fleqn]{mnras}

\usepackage{natbib}
\usepackage{graphicx}
\usepackage{times}
\usepackage{rotate}
\usepackage{hyperref}
\usepackage{url} 
\usepackage[usenames, dvipsnames]{color}
\usepackage{colortbl}
\usepackage[T1]{fontenc}
\usepackage{ae,aecompl}
\usepackage{xcolor}
\usepackage{makecell}
\usepackage{gensymb}
\usepackage{CJKutf8}

\hypersetup{
    linkcolor=blue,citecolor=blue, filecolor=cyan, urlcolor=teal}

\DeclareGraphicsExtensions{.pdf,.png,.jpg}
    
\input{define.tex}

\title[Outskirts of Massive Galaxies]{Reaching for the Edge I: Probing the Outskirts of Massive Galaxies with HSC, DECaLS, SDSS, and Dragonfly}

 \author[J. Li et al.]{
    \href{http://orcid.org/0000-0001-9592-4190}{Jiaxuan Li (李嘉轩)}$^{1,2,3}$\thanks{Email: \href{mailto:jiaxuanl@princeton.edu}{jiaxuanl@princeton.edu}}, 
    \href{https://orcid.org/0000-0003-1385-7591}{Song Huang (黄崧)}$^{3,4}$,
    \href{https://orcid.org/0000-0002-3677-3617}{Alexie Leauthaud}$^{1}$, 
    \href{https://orcid.org/0000-0002-2733-4559}{John Moustakas}$^{5}$, 
    \newauthor 
    \href{https://orcid.org/0000-0002-1841-2252}{Shany Danieli}$^{3,6}$\thanks{NASA Hubble Fellow},
    \href{https://orcid.org/0000-0002-5612-3427}{Jenny E.\ Greene}$^{3}$,
    \href{https://orcid.org/0000-0002-4542-921X}{Roberto Abraham}$^{7}$, \href{https://orcid.org/0000-0002-6100-4852}{Felipe Ardila}$^{1}$, 
    \newauthor 
    \href{https://orcid.org/0000-0002-0332-177X}{Erin Kado-Fong}$^{3}$,  
    \href{https://orcid.org/0000-0002-2406-7344}{Deborah Lokhorst}$^{7}$, 
    \href{https://orcid.org/0000-0003-1666-0962}{Robert Lupton}$^{3}$, 
    Paul Price$^{3}$
    \\
    \\
    $^1$ Department of Astronomy and Astrophysics, University of California Santa Cruz, 1156 High Street, Santa Cruz, CA 95064, USA\\
    $^2$ Department of Astronomy, School of Physics, Peking University, 5 Yiheyuan Road, Haidian District, Beijing 100871, China\\
    $^3$ Department of Astrophysical Sciences, 4 Ivy Lane, Princeton University, Princeton, NJ 08544, USA\\
    $^4$ Department of Astronomy and Tsinghua Center for Astrophysics, Tsinghua University, Beijing 100084, China\\
    $^5$ Department of Physics and Astronomy, Siena College, 515 Loudon Road, Loudonville, NY 12211, USA\\
    $^6$ Department of Astronomy, Yale University, New Haven, CT 06511, USA\\
    $^7$ David A. Dunlap Department of Astronomy and Astrophysics, University of Toronto, 50 St. George Street, Toronto ON, M5S 3H4, Canada\\
}

\date{Accepted XXX. Received YYY; in original form ZZZ}

\pubyear{2021}
\begin{document}

\label{firstpage}

\begin{CJK*}{UTF8}{gbsn}

\pagerange{\pageref{firstpage}--\pageref{lastpage}}
\maketitle

\end{CJK*}

%-------- ABSTRACT  ---------------------
 
\begin{abstract} 
The outer light (stellar halos) of massive galaxies has recently emerged as a possible low scatter tracer of dark matter halo mass. To test the robustness of outer light measurements across different data sets, we compare the surface brightness profiles of massive galaxies using four independent data sets: the Hyper Suprime-Cam survey (HSC), the Dark Energy Camera Legacy Survey (DECaLS), the Sloan Digital Sky Survey (SDSS), and the Dragonfly Wide Field Survey (Dragonfly). We use customized pipelines for HSC and DECaLS to achieve better sky background subtraction.
For galaxies at $z<0.05$, Dragonfly has the best control of systematics, reaching surface brightness levels of $\mu_r \sim 30\ \mathrm{mag/arcsec^2}$. At $0.19<z<0.50$, HSC can reliably recover surface brightness profiles to $\mu_{r} \sim 28.5\ \mathrm{mag/arcsec^2}$ reaching $R=100 - 150$ kpc. DECaLS surface brightness profiles show good agreement with HSC but are noisier at large radii. The median profiles of galaxy ensembles in both HSC and DECaLS reach $R > 200$ kpc without significant bias. At $0.19<z<0.50$, DECaLS and HSC measurements of the stellar mass contained within 100 kpc agree within 0.05 dex. Finally, we use weak gravitational lensing to show that measurements of outer light with DECaLS at $0.19<z<0.50$ show a similar promise as HSC as a low scatter proxy of halo mass. The tests and results from this paper represent an important step forward for accurate measurements of the outer light of massive galaxies and demonstrate that outer light measurements from DECam imaging will be a promising method for finding galaxy clusters for DES and DESI.
\end{abstract}

\begin{keywords}
galaxies: photometry -- galaxies: elliptical and lenticular, cD -- galaxies: structure -- galaxies: formation
\end{keywords}
 
%-------- KEY WORDS  ---------------------

%%%%%%%%%%%%%%%%%%%%%%%%%%%%%%%%%%%%%%%%%%%%%%%%%%%%%%%%%%%%%%%%%%%%%%%%%%%%%%
%     INTRODUCTION
%%%%%%%%%%%%%%%%%%%%%%%%%%%%%%%%%%%%%%%%%%%%%%%%%%%%%%%%%%%%%%%%%%%%%%%%%%%%%%
\section{Introduction}
There is significant interest in understanding exactly how massive galaxies grow with time. The inner regions of massive galaxies have been well studied at low redshifts (e.g., \citealt{Kormendy2009, Lauer2014, MASSIVE2015II, MASSIVE2019XII, MASSIVE2019XIII}). However, accurate measurements of the outer extended envelopes of massive galaxies are not as straightforward. This is because the diffuse stellar halos of massive galaxies are typically low surface brightness (LSB) and are difficult to measure accurately.

The low surface brightness nature of the extended outskirts of massive galaxies makes them extremely sensitive to sky background subtraction. For example, even a small over-subtraction of the background can lead to a sudden truncation of the outer surface brightness profiles and an underestimation of the total luminosity (e.g., \citealt{Blanton2011,Fischer2017,Bernardi2017b}). Moreover, the outer-wing of the point spread function (PSF, e.g. \citealt{Tal2011, DSouza2014,Liu2021PSF}) and the diffuse emission of the dust in the Milky Way (known as the Galactic cirrus; e.g., \citealt{Duc2015,Roman2019}) adds complexity to the analysis of galaxy outskirts.

Furthermore, choices regarding photometric methods can also greatly influence stellar halo studies. When images are not deep enough, stacking techniques are commonly employed (e.g., \citealt{Zibetti2005, Tal2011, DSouza2014,Zhang2019}). Stacking enables statistically significant detection of light in the extreme outskirts of massive galaxies ensembles. However, the results are not free of systematics. When images are deep enough, stellar halos can be studied on an individual galaxy basis \citep[e.g.,][]{Huang2018c}. Here choices involve either using 2-D model fitting or 1-D surface brightness profile analyses. In principle, the 2-D fitting approach has several key advantages (e.g., \citealt{Gonzalez2005,Huang2013a,Miller2021GMM}). Model fitting takes the PSF convolution into account and it is straightforward to integrate the results to get the total luminosity. The 2-D approach can also account for isophotal twists. Yet, in reality, it also faces many challenges, including model bias. For example, simple 2-D models like de Vaucouleurs or single-\sersic{} model cannot reliably describe the light distribution in massive galaxies (e.g., \citealt{Huang2013a, Oh2017}). Therefore, extracting and fitting 1-D surface brightness profiles based on isophotes remains a popular and robust method for exploring the properties of stellar halos \citep{Huang2018c, vanDokkum2014, Iodice2017, Gilhuly2019, Iodice2019}.

Improved measurements of the outer regions of massive galaxies are important for many reasons. First, they are needed for correct estimation of total stellar mass (e.g. \citealt{Bernardi2013, Miller2021}). Second, at large radii, the long dynamical time means that fossil evidence of the assembly history may be preserved (e.g. tidal features from recent mergers, see \citealt{Tal2009, Kado-Fong2018, Smercina2019}). Third, stellar halos show clear dependence on both stellar mass and halo mass, and can therefore provide insight about the galaxy-halo connection (e.g. \citealt{Krick2007, Huang2020ASAP, Montes2019,Huang2021}). Finally, stellar halos are expected to show strong redshift evolution which can be used as a strong constraint on galaxy evolution models (e.g. \citealt{vanDokkum2012, Buitrago2017}).

The goal of this research is to study how far in radius current surveys and techniques can reliably measure the outer regions of massive galaxies.  We focus on four independent data sets that have been used in various analysis of massive galaxies and for studies of LSB features: the Hyper Suprime-Cam Subaru Strategic Program Survey (HSC, \citealt{Aihara2018}), which offers deep images with good seeing but limited sky coverage; the Dark Energy Camera Legacy Survey (DECaLS, \citealt{Dey2019}), with 
wide sky coverage, deeper than SDSS but shallower than HSC; the Dragonfly Wide Field Survey (Dragonfly, \citealt{Abraham2014, Danieli2020, Miller2021}), which is extremely deep and optimized for LSB detection but has poor spatial resolution and small sky coverage currently; and the Sloan Digital Sky Survey (SDSS, \citealt{SDSS-DR7}), which is shallow, has very large sky coverage and has been used in many different studies. We compare the surface brightness profiles extracted from these four surveys as a means of empirically determining the robustness of outer light profile measurements for massive galaxies. We study a low redshift ($z<0.05$) sample using SDSS, Dragonfly, HSC, and DECaLS, and an intermediate redshift sample at $0.19<z<0.50$ using HSC and DECaLS. 

The layout of this paper is as follows. The data and sample selection are described in Section \ref{sec:data&sample}. In Section \ref{sec:methodology}, we describe our image reduction and surface brightness profile measurement methods. The main results of this work are presented in Section \ref{sec:result}. Lessons learned  and future directions are discussed in Section \ref{sec:discussion}. Section \ref{sec:conclusions} presents the summary and conclusions. We adopt a $\Lambda$CDM cosmology with $\Omega_{\rm m}= 0.3$, $\Omega_\Lambda= 0.7$ and $H_0 = 70\ $km s$^{-1}$ Mpc$^{-1}$. We use the AB system \citep{Oke1983} for magnitudes. The stellar mass used in this work is based on a \citet{Chabrier2003} initial mass function.

%%%%%%%%%%%%%%%%%%%%%%%%%%%%%%%%%%%%%%%%%%%%%%%%%%%%%%%%%%%%%%%%%%%%%%%%%%%%%%
%    Data, Sample, Methods
%%%%%%%%%%%%%%%%%%%%%%%%%%%%%%%%%%%%%%%%%%%%%%%%%%%%%%%%%%%%%%%%%%%%%%%%%%%%%%
\section{Data and Sample Selection}\label{sec:data&sample}

We study massive galaxy outskirts using four surveys: HSC, DECaLS, Dragonfly, and SDSS. Dragonfly and SDSS can only be used to study stellar halos at low redshift ($z<0.05$) because of their limited spatial resolution (Dragonfly) and depth (SDSS) . Therefore, we define a low redshift sample that is used to compare the surface brightness profiles among the four different surveys. However, given the current area of the Dragonfly survey, the number of massive galaxies is small (5 galaxies). We also set up a higher redshift sample containing a much larger number of massive galaxies (2171 galaxies), which is used to study the profiles of galaxy ensembles and compare HSC and DECaLS. In this section, we describe our data and sample selection. 

\subsection{Imaging Data}\label{subsec:data}

We first briefly describe the imaging data from the four imaging surveys mentioned above. In this work, we use the $r$-band data for comparing surface brightness profiles. However, the effective $r$-band transmission curves vary among surveys. We perform filter correction using the method described in Appendix \ref{subsec:filt-corr}.

\subsubsection{Hyper Suprime-Cam Survey}
\label{subsec:hsc}

%HSC
The Hyper Suprime-Cam Subaru Strategic Program Survey (\citealt{Aihara2018}; hereafter `HSC survey')\footnote{\url{https://hsc-release.mtk.nao.ac.jp/doc/}} is based on the Hyper Suprime-Cam \citep{Miyazaki2012, Miyazaki2018}, which is a prime-focus camera mounted on the 8.2-m Subaru telescope. The \texttt{WIDE} layer of the HSC survey aims to provide deep images in five broad-bands ($grizy$) over 1000 deg$^2$. The depth of the \texttt{WIDE} layer is $g=26.6$ mag, $r=26.2$ mag and $i=26.2$ mag ($5\sigma$ point source). The responses of the HSC $griz$-band filters are similar to those of SDSS \citep{Kawanomoto2018}. HSC has a pixel scale of $0.168\arcsec$ per pixel and a typical seeing of $0.8\arcsec$ in the $r$-band, giving it the best resolution among current wide-field sky surveys. The HSC footprint overlaps with many spectroscopic surveys, such as the Sloan Digital Sky Survey (SDSS) legacy survey \citep{SDSS-DR16} and the Baryon Oscillation Spectroscopic Survey (BOSS, \citealt{Dawson2013}), the Galaxy and Mass Assembly (GAMA, \citealt{Driver2011})\footnote{\url{http://www.gama-survey.org/}} survey, and the Visible Multi-Object Spectrograph Very Large Telescope Deep Survey (VVDS, \citealt{VVDS2005})\footnote{\url{https://cesam.lam.fr/vvds/}}. 

HSC data are processed using \code{hscPipe}\footnote{\url{https://hsc.mtk.nao.ac.jp/pipedoc_e/}} \citep{Bosch2018}, which is a modified version of the Large Synoptic Survey Telescope (LSST) pipeline \citep{LSST-pipeline}\footnote{\url{https://pipelines.lsst.io/}}. \code{hscPipe} performs forced multi-band photometry of objects using \code{cModel}, which fits each object with a combination of a de Vaucouleurs and an exponential components after taking PSF convolution into account. 

%The \code{cModel} photometry is robust and accurate down to $i \sim 25.0$ mag \citep{Huang2018PASJ} but it tends to underestimate the flux of bright object (e.g., $i<20.0$ mag). Moreover, the intrinsic limitation of the \code{cModel} assumption makes it not suitable to acount for the extended outskirts of massive early-type galaxies (ETGs) and significantly underestimates the total fluxes of them \citep{Huang2020ASAP}. 

Sky background subtraction is a prominent difficulty for the photometry of massive galaxies. In the HSC data release \code{S18A} (i.e., Public Data Release 2, \code{PDR2}, \citealt{HSC-PDR2}), \code{hscPipe 6.7} implemented a full focal plane sky subtraction algorithm to overcome the over-subtracted local background around bright objects. It also provided sky objects for calibrating the sky background (see Section \ref{sec:bkg-mock-test} and Appendix \ref{appendix:mock-test} for details). Both make it more suitable for studying low surface brightness universe. HSC \code{S18A} also provides bitmasks indicating bad pixels, cosmic rays, edges of CCDs and pixels with source detection, helping us generate image masks when extracting surface brightness profiles (Section \ref{sec:hsc_methods}). In this paper, we use the \code{WIDE} layer data from \code{S18A} (\code{PDR2}). It covers $\sim 300$ deg$^2$ in all five bands.

\subsubsection{DECam Legacy Survey}
\label{subsec:decals}

%DECaLS
The Dark Energy Camera Legacy Survey (DECaLS)\footnote{\url{https://www.legacysurvey.org/decamls/}} is a public imaging survey using the Dark Energy Camera (DECam, \citealt{DECam2008}) on the 4-m Blanco telescope. It covers a 9000 deg$^2$ region within $-20^{\circ}\leqslant\mathrm{DEC}\leqslant32^{\circ}$ in $grz$ bands with a 5$\sigma$ point source detection limit of $g=24.0$, $r=23.4$ and $z=22.5$ mag. 
DECaLS is also an important part of the Dark Energy Spectroscopic Instrument (DESI) Legacy Imaging Surveys \citep{Dey2019}.
The DECaLS data are reduced and analyzed by a specialized photometric pipeline \code{legacypipe}\footnote{\url{https://github.com/legacysurvey/legacypipe}}.
The pipeline relies on the probabilistic source detection and measurement tool \tractor{}\footnote{\url{https://github.com/dstndstn/tractor}} for photometric analysis.
Please see \citet{Dey2019} for details about the data reduction.
In this paper, we use the most up-to-date data release (DR9\footnote{\url{http://legacysurvey.org/dr9/description/}}) with customized, CCD-level sky background-subtraction. We describe this procedure in Section \ref{sec:decals_methods}.

\subsubsection{Dragonfly Wide Field Survey}
\label{subsec:dragonfly}

%Dragonfly
The Dragonfly Telephoto Array (`Dragonfly' for short, \citealt{Abraham2014, Danieli2020}) is an imaging system consisting of 48 Canon 400 mm $f/2.8$ IS II telephoto lenses, equivalent to a 1\,m aperture, $f/0.4$ refractor. The system is designed and optimized for detecting LSB features. Each lens is mounted to a CCD camera, providing a $2.6\degree \times 1.9\degree$ instantaneous field of view with a $2.8\arcsec$ native pixel scale, and a seeing with $\approx 5\arcsec$ full-width-half-maximum (FWHM). Half of the lenses are equipped with SDSS-$g$ filters and the other half with SDSS-$r$ filters. The all-refractive design and `sub-wavelength nanostructure' anti-reflection coatings of the Canon telephoto lenses ensure that Dragonfly's PSF is very well-controlled with low power distributed in far wings \citep{Liu2021PSF}. Dragonfly is capable of detecting surface brightness down to $\mu_g = 32\ \mathrm{mag/argsec}^2$ in 1-D surface brightness profiles \citep{vanDokkum2014}. In this paper we use data from the Dragonfly Wide Field Survey (\citealt{Danieli2020}). The footprint of the survey covers $330\ \mathrm{deg}^2$ and overlaps with Stripe 82 and the GAMA fields to maximize its scientific potential with other multi-wavelength data sets. The methods used for Dragonfly data reduction are further described in Section \ref{sec:dragonfly_methods}.

\subsubsection{Sloan Digital Sky Survey}
\label{subsec:sdss}

%SDSS
We also use images from the Sloan Digital Sky Survey \citep{Gunn2006, SDSS-DR7}. SDSS provides $ugriz$ images for 14,500 deg$^2$ with a 5$\sigma$ point source detection limit of $g=23.13$, $r=22.70$, $i=22.20$, and $z=20.71$ mag. We generate coadds using \code{SWarp}\footnote{\url{http://www.astromatic.net/software/swarp}} \citep{Swarp} based on SDSS DR7 images \citep{SDSS-DR7} and using the SDSS image mosaic tool\footnote{\url{https://dr12.sdss.org/mosaics}}. 
SDSS is the shallowest imaging survey used in this work and is not a good choice for studying the LSB universe 
except for the $2$ mag deeper Stripe 82 region (e.g., \citealt{Fliri2016, Roman2019}).
It is also well-known that SDSS images suffer from over-subtraction around bright objects (e.g., \citealt{Blanton2011, Bernardi2013}).
We do not attempt to correct the over-subtraction in this work since we only use the SDSS images as a reference for the low redshift sample.

\subsection{Low Redshift Galaxy Sample}\label{subsec:low-z-sample}

% Low-z postage
\begin{figure*}
	%\hskip -5mm
	\vbox{ 
		%\vskip -10mm
		\centering
		\includegraphics[width=1\linewidth]{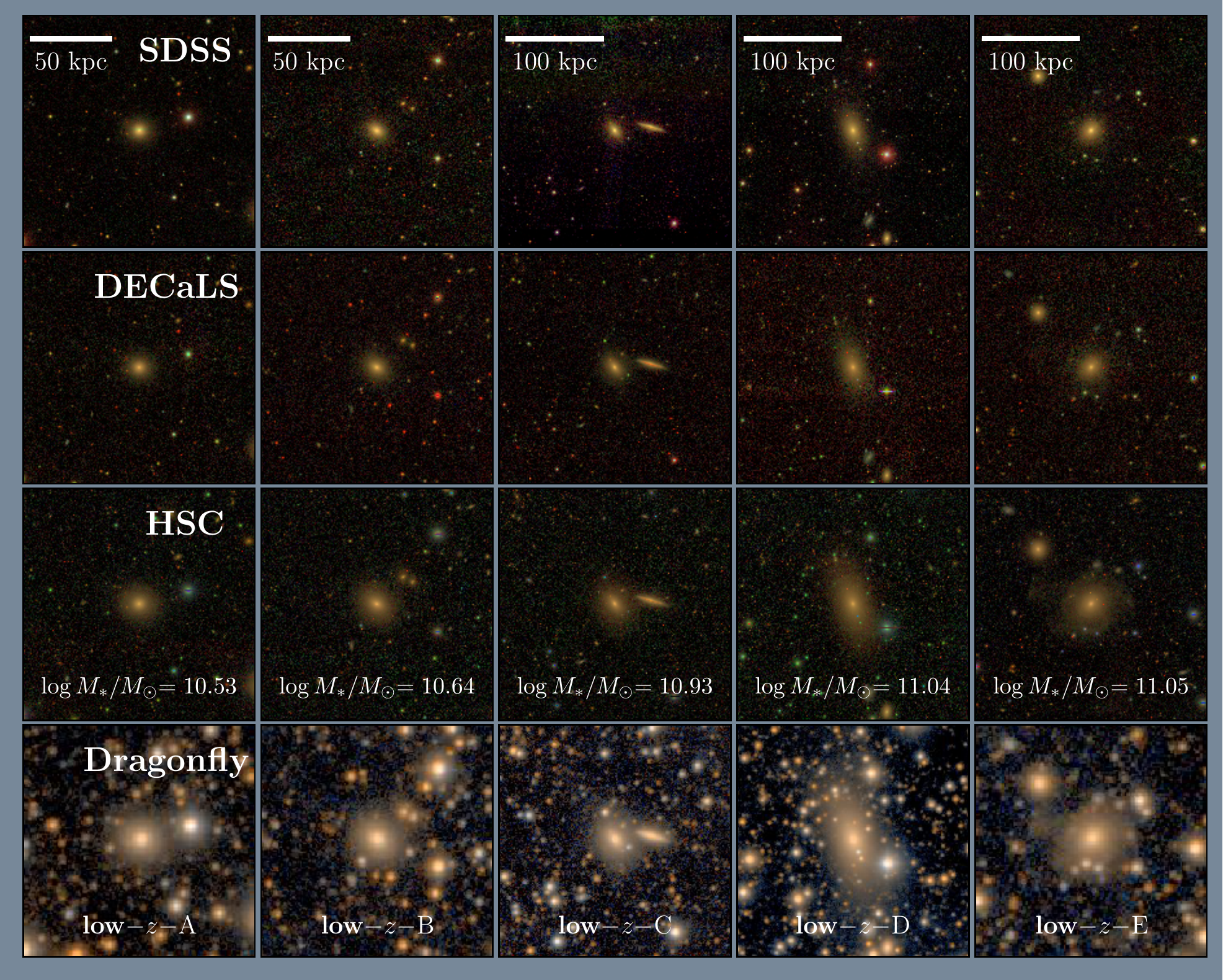}
	}
	\caption{Tri-color images of five massive galaxies in the low-$z$ sample (Table \ref{tab:low-z-table}). Images from top to bottom correspond to SDSS, DECaLS, HSC, and Dragonfly. The two galaxies on the left have an image size of 150 kpc $\times$ 150 kpc, other galaxies have an image size of 250 kpc $\times$ 250 kpc.} %The HSC point source detection is $\sim2.6$ mag deeper than DECaLS and $\sim3.5$ mag deeper than SDSS. Of the four surveys, Dragonfly has the worst spatial resolution and seeing but an excellent ability to map the low surface brightness features of low-$z$ galaxies.
	\label{fig:low-z-postage}
\end{figure*}

One of the primary goals of this work is to evaluate systematic issues that arise when measuring the low surface brightness outskirt of massive galaxies. 
Such issues often become more prominent for very low-redshift early-type galaxies (ETGs) that have large angular sizes and bright magnitudes.
For this purpose, we cross-match the four surveys introduced above in their overlapping regions (including some of the GAMA fields and part of the Stripe 82 region).
We first select galaxies with $z\leqslant0.04$ and $\log (M_{*}^{\mathrm{GAMA}}/M_\odot) > 10.5$ using the GAMA stellar mass catalog \citep{GAMA-mass} in the overlapping regions as an initial sample.
After applying the HSC bright star masks \citep{Coupon2018} and reviewing the HSC images, we exclude galaxies that are saturated in $r$-band or contaminated by nearby saturated stars.
We only consider galaxies with early-type morphologies and extended outskirts.
We then select the top five most massive galaxies as our low redshift galaxy sample (the `low-$z$ sample'). Their stellar masses are in the range of $10^{10.5}M_\odot$ to $10^{11.1}M_\odot$, which is lower than the galaxies from the intermediate redshift sample (see Section \ref{subsec:mid-z-sample}) because of the smaller volume probed.
Basic information about the sample is listed in Table \ref{tab:low-z-table}.  
Figure \ref{fig:low-z-postage} shows tri-color images of our low-$z$ sample. The tri-color images for HSC, DECaLS, and SDSS are generated using three bands data ($gri$ for HSC and SDSS, $grz$ for DECaLS) according to the color scheme described in \citet{Lupton2004}. Since Dragonfly has only $g$- and $r$-band data, we make synthetic color images using $g$-band intensity $I(g_{\mathrm{DF}})$ as the blue channel, $r$-band intensity $I(r_{\mathrm{DF}})$ as the red channel, and $0.5\,I(g_{\mathrm{DF}}) + 0.5\,I(r_{\mathrm{DF}})$ as the green channel.

%The DWFS footprint covers Stripe82 and some of the GAMA fields. The HSC \code{S18A} \texttt{WIDE} layer 
%footprint overlaps with Dragonfly in the G09, G12, and G15 GAMA fields. 
%We cross-match the GAMA catalogue \citep{GAMA-mass} with both HSC and Dragonfly data, then select ETGs with $z\leqslant0.04$ and $\log (M_{*}^{\mathrm{GAMA}}/M_\odot) > 10.5$ as an initial sample. For HSC, a major concern is that very nearby massive galaxies are typically saturated in $r$- or $i$-band. Since ETGs are typically brighter in $i$-band and thus are prone to be saturated, we focus only on $r$-band data in this paper (in addition, DECaLS does not have $i$-band). We apply the bright star mask and also exclude $r$-band saturated galaxies in the initial sample. 

% Low-z information
\begin{table}
    \centering
    \begin{tabular}{ccccc}
    \hline
      Name & R.A. & Dec. & $z$ & $\log M_*/M_\odot$ \\
      \hline
      low-$z$-A & 175.02673 & -0.83771 & 0.020 & 10.53 \\
      low-$z$-B & 177.40144 & -1.45549 & 0.019 & 10.64 \\
      low-$z$-C & 177.37548 & -1.08661 & 0.019 & 10.93 \\
      low-$z$-D & 140.00904 & 1.03827 & 0.017 & 11.04 \\
      low-$z$-E & 222.48908 & 0.55829 & 0.040 & 11.05 \\
     \hline
    \end{tabular}
    \caption{Name, position, redshift, and stellar mass of galaxies in the low-$z$ sample. The stellar mass values are retrieved from the GAMA catalog \citep{GAMA-mass}.}
    \label{tab:low-z-table} 
\end{table}

\subsection{Intermediate Redshift Galaxy Sample}\label{subsec:mid-z-sample}

% Mid-z distribution
\begin{figure*}
    \vbox{ 
		%\vskip -10mm
		\centering
		\includegraphics[width=0.8\linewidth]{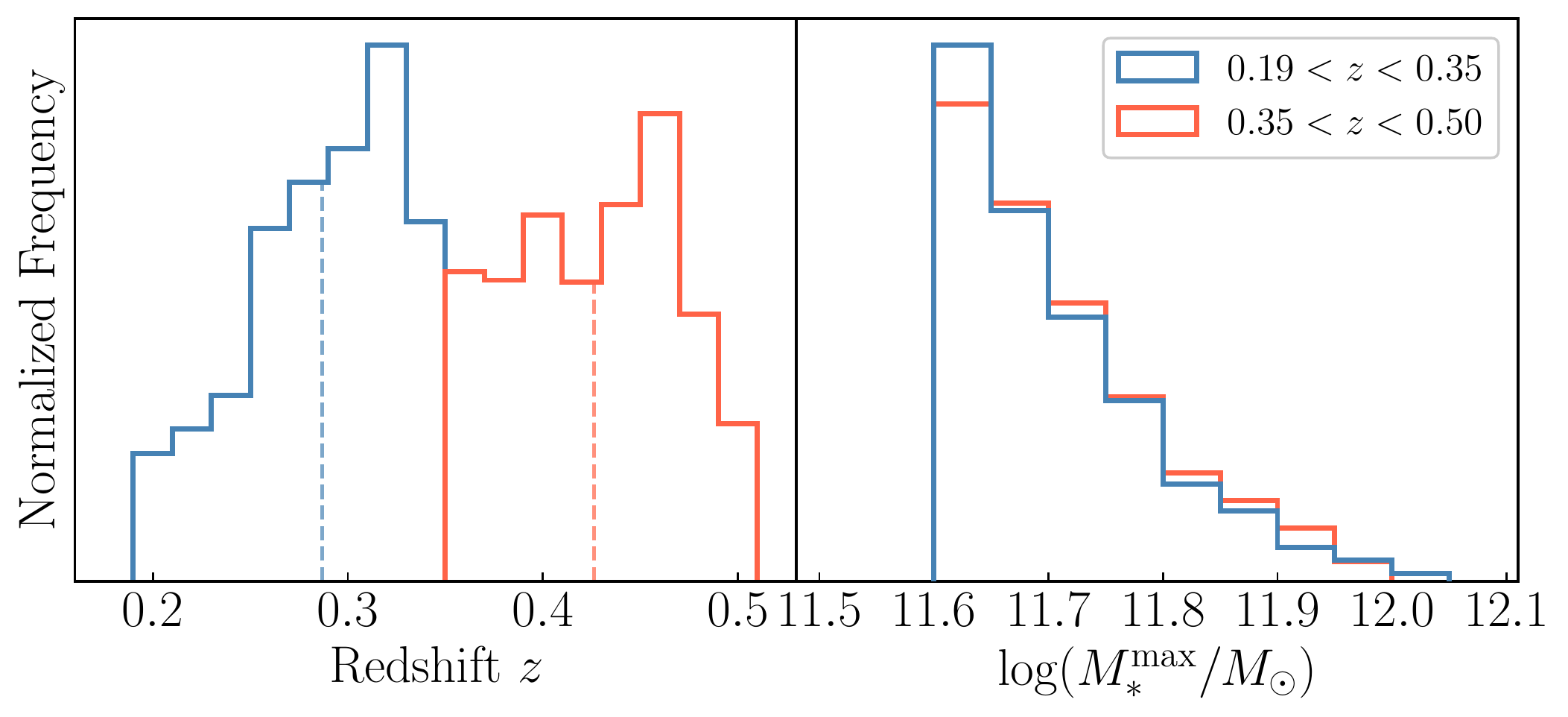}
	}
	\caption{Spectroscopic redshift (left) and stellar mass distributions (right) of the intermediate-$z$ sample. The blue and red solid lines correspond to the two redshift bins. In the left panel, the mean redshifts of each bin are marked by dashed vertical lines. The stellar mass distributions of the two samples are similar.}
	\label{fig:mid-z-distribution}
\end{figure*}

Another motivation of this work is to compare the surface brightness limits of HSC and DECaLS images for massive galaxies.
This comparison can help us determine whether it will be possible to take advantage of the larger sky coverage of DECaLS (including spectroscopic redshifts from DESI) to study the outer light of massive galaxies.

We construct an intermediate redshift sample very similar to the sample used in \citet{Huang2020ASAP}. We select galaxies in the HSC footprint with spectroscopic redshifts $0.19 < z < 0.50$ and stellar mass $\log (M_{*}^{\mathrm{max}}/M_{\odot}) > 11.6$, where $M_{*}^{\mathrm{max}}$ is the maximum 1-D stellar mass derived from the surface brightness profile using the HSC data release \code{S16A} (see \citealt{Huang2018c} for details). 
To measure $M_{*}^{\mathrm{max}}$, we integrate the surface brightness profile to the radius where the median intensity is consistent with the standard deviation of the sky background. We derived the average mass-to-light ratio and $k$-corrections using the broad-band spectral energy distribution fitting code \code{iSEDfit}\footnote{\url{https://github.com/moustakas/iSEDfit}} \citep{Moustakas2013,Moustakas2017}.
For massive galaxies, the maximum 1-D stellar mass is typically larger than the \code{cModel} stellar mass by $0.2-0.3$ dex. 
Compared to the fixed aperture stellar mass such as $M_{*}^{100\ \rm kpc}$ used in earlier work (e.g. \citealt{Huang2018c}), $M_{*}^{\rm max}$ adds up to $\rm 0.05$ dex of additional stellar mass and is closer to the true ``total'' stellar mass of the galaxy. However, the choice of stellar mass definition for the initial selection has no impact on the results of this work.
Unless noted otherwise, we will use $M_{*}^{\rm max}$ as the default stellar mass for the intermediate-$z$ sample.

The redshift range used here ensures that the sky background subtraction and cosmological dimming are not major issues for studying galaxy outskirts and the galaxies are not saturated at the center of the $r$-band images.
We can also largely ignore the redshift evolution of galaxy structure across this redshift range which is equivalent to a $\sim 2.5$ Gyr of time.

\begin{figure*}
	%\hskip -5mm
	\vbox{ 
		%\vskip -10mm
		\centering
		\includegraphics[width=1\linewidth]{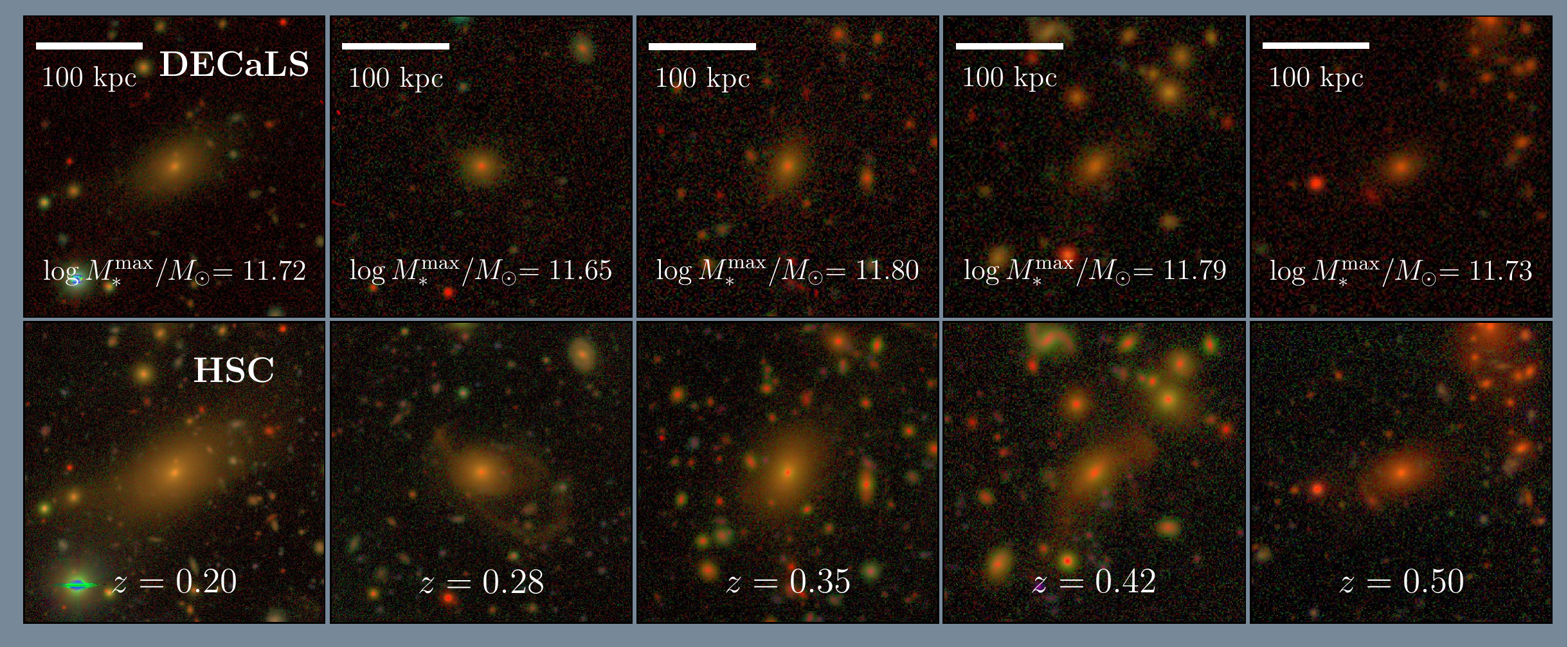}
	}
	\caption{Tri-color images for five randomly selected galaxies in the intermediate-$z$ sample from DECaLS and HSC. Images are generated following the same format as in Figure \ref{fig:low-z-postage}. Images are 300 kpc $\times$ 300 kpc, with maximum 1-D stellar mass labelled. Deep images reveal extended outskirts and evidence of recent mergers.}
	\label{fig:mid-z-postage}
\end{figure*}

For each galaxy, we detect neighboring objects using \code{Source Extractor} \citep{Bertin1996,Barbary2016}. We remove galaxies with very close satellites or strong contamination by bright stars and artifacts. After filtering galaxies with the \code{S18A} bright star mask and full-color full-depth mask \citep{Coupon2018,HSC-PDR2}, we obtain 2171 galaxies in the intermediate redshift sample (`intermediate-$z$ sample'). This sample has reliable spectroscopic redshift measurements from BOSS, SDSS, or GAMA. The distributions of spectroscopic redshifts and stellar masses of the intermediate-$z$ sample are shown in Figure \ref{fig:mid-z-distribution}. Five randomly selected galaxies from the intermediate-$z$ sample are shown in Figure \ref{fig:mid-z-postage}.

Whereas HSC images have sufficient resolution and depth to study $z>0.4$ galaxies, it is potentially more challenging for DECaLS to do so. 
Therefore, we separate the intermediate-$z$ sample into two redshift bins: $0.19 < z < 0.35$ (870 galaxies, with mean redshift of $\overline{z}=0.29$) and $0.35 < z < 0.50$ (1301 galaxies, with mean redshift of $\overline{z}=0.43$). 
We neglect redshift-dependent effects within each redshift bin. We compare the two redshift bins as a test for redshift dependent effects. Figure \ref{fig:mid-z-distribution} shows that the two redshift bins have very similar stellar mass distributions. 

% Mid-z postage

\section{Methodology}\label{sec:methodology}
In this section, we present the methodology used to extract 1-D profiles using different survey data, including sky background tests and corrections. As mentioned before, filter correction must be applied to obtain consistent photometry among surveys. We derive a filter correction scheme using the Gunn-Stryker stellar spectra atlas, presented in Appendix \ref{subsec:filt-corr}. We apply the filter corrections to the surface brightness profiles and present the results in Section \ref{sec:result}.

\subsection{HSC Methodology}\label{sec:hsc_methods}
In this subsection, we describe our methodology for extracting 1-D surface brightness profiles using HSC \code{S18A} data. We also test the sky background in HSC \code{S18A} and proposed an empirical correction scheme. Our methods are validated using a variety of tests with synthetic galaxies.

\subsubsection{Measurements of surface brightness profiles}\label{sec:SBP-method}

We use the elliptical isophote fitting program {\ellipse} \citep{Jedrzejewski1987} in IRAF (Image Reduction and Analysis Facility, \citealt{IRAF1986,IRAF1993}) to measure surface brightness profiles. We follow the methodology used in \citet{Huang2018c} with some modifications. We summarize the main steps here and highlight updates. We refer the reader to \citet{Huang2018c} for further details.

We download $r$-band co-added cutout images with sizes corresponding to $1000\ \mathrm{pixels} \times 1000\ \mathrm{pixels}$ for the intermediate-$z$ sample. This corresponds to 260 kpc at $z=0.19$ and 510 kpc at $z=0.50$ in radius. We use a variable cutout sized for the low-$z$ sample. For \code{low-z-A} and \code{low-z-B}, we generate 150 kpc $\times$ 150 kpc cutouts. For the other three, we adopt a 250 kpc $\times$ 250 kpc size due to their large effective radii. The cutout images are large enough for us to measure the surface brightness profile down to the limiting imaging depth and to estimate the local sky background.

We detect bright objects and mask them out using {\sep} \citep{Barbary2016}, a Python implementation of \code{Source Extractor} \citep{Bertin1996}. Specifically, we extract objects above a 4.5$\sigma$ threshold and convolve the segmentation map with a 15-pixel radius Gaussian kernel to increase the mask size. We aggressively mask out very bright objects ($r_{\mathrm{mag}} < 18.0$) by making elliptical masks with the shape and size of each neighbouring galaxy from \code{SExtractor} but enlarged the size by a factor of 10. Considering that \code{SExtractor} segmentation map do not always correctly outline irregular extended objects (such as tidal tails), we try to incorporate the HSC detection masks with the \code{SExtractor} mask. In order to protect the central galaxy from being masked out, we make a hybrid mask: we use only the \code{SExtractor} mask within a certain radius and use the direct combination of the \code{SExtractor} mask and the HSC detection mask outside of that radius. This transition radius is $6R_e$ for galaxies at $z>0.35$, and $10R_e$ for galaxies at $z<0.35$, where $R_e$ is the effective radius. Finally, we combine this mask with the HSC bright star bitmask\footnote{\url{https://hsc-release.mtk.nao.ac.jp/doc/index.php/bright-star-masks-2/}} to shield the light from nearby bright stars and galaxies. 
In this way, almost all bright objects, saturation tails, and tidal disruption debris are masked out except our target object. To test the robustness of our masking method, we adjusted the key parameters (detection threshold, Gaussian kernel radius, etc.) several times, and found negligible changes to our measured profiles. We also compare this binary masking method with a more sophisticated 2-D modeling method in Appendix \ref{subsec:2Dmodel}, where we find no significant difference between the two methods in terms of extracting of the 1-D profiles for the intermediate-$z$ galaxies.

Next, we run {\ellipse} on the cutout image with the mask described above. We first fit isophotes with a fixed center but allow the shape (i.e., ellipticity and position angle) to vary freely. We estimate the mean ellipticity and position angle within 20 kpc $\leqslant R \leqslant$ 50 kpc for the intermediate-$z$ sample, where $R$ is the radial distance along the semi-major axis of elliptical isophotes. For galaxies in the low-$z$ sample, we set this interval to 20 kpc $\leqslant R \leqslant$ 30 kpc. Then we measure the surface brightness profile along the semi-major axis with a fixed center and shape. The mean ellipticity and position angle in the radius interval from the first run are taken as the fixed isophote shape in the second run.

We extract surface brightness profiles along the semi-major axis using a 0.1 dex logarithmic step size. 
Within each elliptical annulus that defines the isophote, we perform sigma-clipping with a $3\sigma$ upper limit and $2.5\sigma$ lower limit three times to reduce the impact of unmasked or undetected neighbors. After sigma-clipping, we take the median value of pixels within each annulus. These choices are made to ensure that the final surface brightness profiles are not affected by any nearby compact objects that are not identified and masked out. In this work, we present all the surface brightness profiles with an $R^{1/4}$ scaling to emphasize the outskirts. We also show the corresponding $R$ in linear scale on the top axis.

For this paper, rather than directly trusting the sky subtraction from \code{hscPipe}, we manually apply empirical corrections to the background subtraction (see Section \ref{sec:bkg-mock-test} for further details). Background corrections are applied directly to profiles, instead of at the image level. We do not find any notable difference between the two approaches. 
{\ellipse} also gives the measurement error of the intensity along the semi-major axis, but the errors estimated by {\ellipse} are usually underestimated because they do not take the uncertainty from sky subtraction into account. We do not add sky subtraction uncertainty into single profiles. The variance between profiles visually highlights the sky uncertainty (e.g., in Figure \ref{fig:mid-z-linear}). 

Median surface brightness profiles of a subset of galaxies are useful when comparing two different data sets. In this paper, the median profile is first calculated in intensity units, then converted to a surface brightness in units of $\mathrm{mag/arcsec^2}$ if required. Individual profiles are interpolated and evaluated on an even grid between $[R / \rm{kpc}]^{1/4} = 1$ and $[R / \rm{kpc}]^{1/4} = 5.5$ with step size $[\Delta R / \rm{kpc}]^{1/4} = 0.05$. Changing the interpolation grid does not alter any of our results. We do not exclude any negative intensity values when calculating median profiles. Using magnitudes instead of linear intensity values can lead to artifacts, as discussed in Section \ref{subsec:linear-profile}. 

\subsubsection{Tests on the Robustness of HSC Profiles to Background Subtraction}\label{sec:bkg-mock-test}

% Mock test comparison
\begin{figure*}
	%\hskip -5mm
	\vbox{ 
		%\vskip -10mm
		\centering
		\includegraphics[width=1\linewidth]{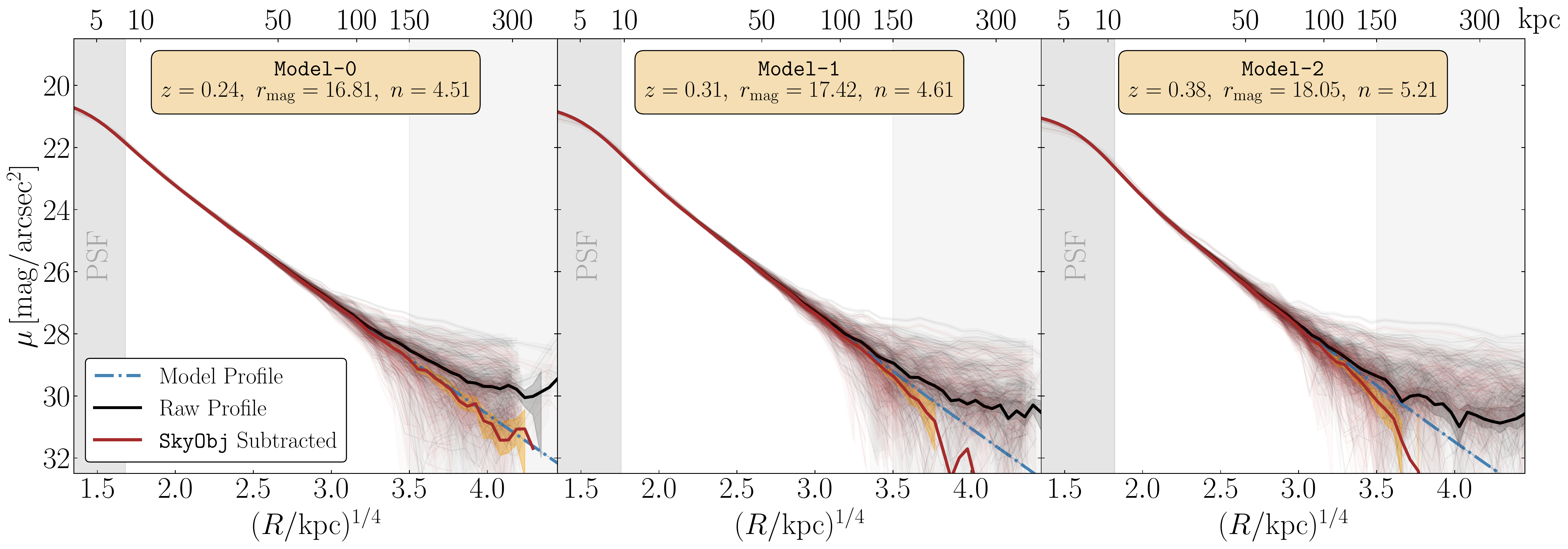}
	}
	\caption{Tests on the accuracy of HSC sky estimation using {\skyobjs} with synthetic galaxies. We test three different galaxy models (see Table \ref{tab:fake-gal-info}). The blue dash-dotted line indicates the fiducial model profile. Thin grey lines are individual realizations of synthetic galaxies. The thick line is the median profile of 60 mock galaxies. The value of the sky is estimated using the mean value of {\skyobjs} with $8.4\arcsec$ apertures which are selected within an annulus of $1\arcmin$ and $4\arcmin$. The sky subtracted profile is shown by the red line. These tests demonstrate two things. First, a small sky residual exists in HSC \code{S18A} images. Second, the use of {\skyobjs} allows us to reliably estimate the sky background and to extract profiles to $\sim 150$ kpc and $\mu = 29.0 - 29.5\ \mathrm{mag/arcsec}^2$. }
	\label{fig:mock-test-compare}
\end{figure*}

In measuring the surface brightness profile of massive galaxies, sky subtraction plays a critical role (e.g., \citealt{Huang2013a, Huang2018c}). In data releases earlier than \code{S18A} (\code{PDR2}), \code{hscPipe} measured the sky by fitting a map of `super-pixels' (256 pixels $\times$ 256 pixels) with a sixth-order 2-D Chebyshev polynomial in each exposure. However, in this method, the sky model was discontinuous at the boundaries of CCDs and could easily be biased by nearby bright objects. Therefore, the old \code{hscPipe} was found to over-subtract the sky background around bright objects, leaving `dark rings' around bright galaxies. This issue limited studies of the LSB regime and is the reason why \citet{Huang2018c} had to apply an empirical local background correction to their 1-D surface brightness profile and only considered profile within 100 kpc as reliable. In data release \code{S18A}, \code{hscPipe} implements a global sky subtraction method that significantly reduce the local over-subtraction issue around bright objects. This method measures an empirical background model over the entire focal plane and further corrects the residual background pattern caused by differences of detector characteristic and spatial variation of filter response function \citep{HSC-PDR2}. While the \code{S18A} data have much better sky subtraction and preserve LSB features very well (see Figure 5 in \citealt{HSC-PDR2}), they could still have residual background on the coadd images that affects the measurement of surface brightness profiles.
In \code{S18A}, \texttt{hscPipe} also performs measurements for artificial ``objects'' in empty regions in which no object is detected and no surrounding objects are detected in any of the five bands. Then the fluxes within these voids are measured on the coadd level with different aperture sizes. These empty regions are called `Sky Objects' ({\skyobjs}). 
The aperture fluxes of these \skyobjs{} are good proxies of the local sky background although the measured background level may depend on the choice of the aperture size.

We conduct a series of tests using mock intermediate-$z$ galaxies to check the quality of sky subtraction in HSC \code{S18A} and verify the robustness of our methodology using the following procedures.

First, we make $r$-band cutouts with centers located at the positions of randomly selected {\skyobjs}, which provide convenient locations to inject mock objects and generate realistic images. Then we generate mock intermediate-$z$ galaxies using {\galsim}\footnote{\url{https://github.com/GalSim-developers/GalSim}} \citep{GalSim2015} and add them to these images. We retrieve the PSF model generated by \code{hscPipe} at the center of each cutout and use it to convolve the \galsim{} model.

Although recent studies have found that ETGs are not well described by single-{\sersic} models (e.g. \citealt{Huang2013a, Oh2017}), for our fake galaxy tests, we opt for simple single-{\sersic} fitting with parameters drawn from the GAMA survey \citep{GAMA2012}. We randomly choose three different models with relatively large {\sersic} indices ($n>4$) to mimic the extended outskirts of real massive ETGs. Model choices for our fake galaxy tests are listed in Table \ref{tab:fake-gal-info}. In total, we generate mock images for 60 randomly selected positions with three different models located at different redshifts.

% Fake galaxy info
\begin{table}
    \centering
    \begin{tabular}{ccccccc}
    \hline
      Name   & $z$ & $r_{\mathrm{mag}}$ & {\sersic} $n$ & \makecell{$R_e$ \\ (arcsec)} & $e$ & P.A. \\
      \hline
      \texttt{Model-0} & 0.240 & 16.81 & 4.514 & 5.86 & 0.166 & -57.17\\
      \texttt{Model-1} & 0.309 & 17.42 & 4.613 & 3.95 & 0.178 & -16.18\\
      \texttt{Model-2} & 0.383 & 18.05 & 5.211 & 2.67 & 0.226 & 40.90\\
     \hline
    \end{tabular}
    \caption{Three different galaxy models used in mock tests. Models parameters are drawn from single-{\sersic} modeling from the GAMA survey \citep{GAMA2012}. Our fake galaxies span a large range of redshift, $r$-band magnitude, {\sersic} index and shape.}
    \label{tab:fake-gal-info} 
\end{table}

We measure 1-D surface brightness profiles for mock galaxies using the method described in Section \ref{sec:SBP-method}. The resulting profiles are shown in Figure \ref{fig:mock-test-compare} as dim black lines. The thick black line is the median profile of all mock galaxies, and blue dash-dotted lines are fiducial model profiles. The flattened outskirt of the median profile indicates that the HSC $r$-band sky for \code{S18A} is under-subtracted, instead of over-subtracted as seen in previous data releases \citep{Huang2018c}. We also found that the overall residual level is filter-dependent and aperture-dependent (see Appendix \ref{appendix:mock-test} and Figure \ref{fig:skyobj_stats}).

Residual background levels significantly affect measurements of LSB outer profiles. We therefore correct for this effect using the mean value of nearby {\skyobjs}. This largely mitigates the small under-subtraction issue in the $r$-band. The mean correction for the sky background is typically in the range of 0.002 -- 0.008 $\mu\mathrm{Jy/arcsec}^2$ (0.001 -- 0.004 counts/pixel). Based on our mock tests, measuring the sky using $8.4\arcsec$ aperture {\skyobjs} matched within $[1\arcmin, 4\arcmin]$ yields the most robust results. This is shown in Figure \ref{fig:mock-test-compare} where the red line displays the median profile using this sky correction scheme based on {\skyobjs}. From Figure \ref{fig:mock-test-aperture}, we find that a $8.4\arcsec$ aperture size works well for bright low-$z$ galaxies, whereas a $5.7\arcsec$ aperture size is better for fainter intermediate-$z$ galaxies. However, Section \ref{subsec:linear-profile} and Appendix \ref{appendix:mock-test} show that a $8.4\arcsec$ aperture is also appropriate for the intermediate-$z$ sample. We use a $8.4\arcsec$ aperture for both samples in the remainder of this paper. Appendix \ref{appendix:mock-test} includes further details on mock tests and {\skyobjs}. 

For low-$z$ bright galaxies, the sky objects around these galaxies are rare compared with intermediate-$z$ galaxies and are typically far away from the over-subtracted ring. Hence, \skyobjs{} are no longer good proxies of the over-subtracted sky for these objects. Thus, we measure the sky background locally and correct for over-subtraction in the  1-D profiles using the following method. We first aggressively mask out every object in the image by combining a high-threshold ($\sim 5\sigma$) source extraction and a low-threshold ($\sim 2\sigma$) source extraction. Then we bin the masked image with a $6\arcsec\times6\arcsec$ box (which is larger than the local fluctuation of the image), and we find that the intensity distribution of these `super-pixels' can be well described by a Gaussian profile. We then adopt the median of the distribution as the local sky background value. We then correct for this local sky value for all low-$z$ galaxies.  %We also estimate the standard error of the median (referred to as the \emph{background uncertainty}) to be $\sigma/(f\sqrt{N})$, where $\sigma$ is the standard deviation of super-pixel values, $N$ is the number of super-pixels and $f$ is the binning factor. One can also calculate this background uncertainty by bootstrap resampling, which gives almost the same results as $\sigma/(f\sqrt{N})$. This uncertainty is used to construct the upper and lower-limit profiles in Section \ref{subsec:lowz-result}.

\subsection{DECaLS Methodology}
\label{sec:decals_methods}

Here we briefly introduce the procedure for extracting 1-D surface brightness profiles of massive galaxies using DECaLS images.
The DECaLS public data release (here, DR9) includes tools to generate coadds using all the single exposure images in a given band for positions within the survey footprint. The algorithm used to model and subtract the sky background in DR9, however, systematically over-subtracts the local background around bright objects, hence affecting the LSB outskirts of massive galaxies in our sample (see \citealt{Dey2019} and J. Moustakas et al., in preparation). To mitigate this issue, we use a custom algorithm to subtract the small-scale background\footnote{The large-scale, inter-CCD sky background, including the pupil ghost and other large-scale additive background terms, have already been subtracted from the data at this point in the pipeline, leaving only the small-scale, intra-CCD background (see \citealt{Dey2019} for details).} from each individual CCD before building the DECaLS mosaics and performing our photometric measurements.

First, we determine which CCDs overlap an area of the sky in fixed physical (comoving) coordinates based on the central coordinates and redshift of each galaxy.  For the intermediate-$z$ sample we use a $500$~kpc$\times500$~kpc area, while for the low-$z$ sample we use $300$~kpc$\times300$~kpc.\footnote{For reference, the typical angular diameter of these cutouts is $\approx1.5$~arcmin ($\approx370$~pixels) for the intermediate-$z$ sample and $\approx20$~arcmin ($\approx5000$~pixels) for the low-$z$ sample, while the area of a single DECam CCD is roughly $9 \times 18\ \mathrm{arcmin}^{2}$.} Next, we bring all the input imaging in a given filter to a common (additive) pedestal by solving the linear equation $\mathbf{A}\mathbf{x}=\mathbf{b}$ for the full set of $N$ images and $M$ overlapping images. In this equation, $\mathbf{A}$ is an $M \times N$ matrix of the summed variances of overlapping CCD pixels, $\mathbf{x}$ is an $N$-length vector of the additive offsets we must apply to each image (in calibrated flux units), and $\mathbf{b}$ is an $M$-length vector of the summed inverse-variance weighted difference between the pixels of overlapping images. We solve this equation using a standard linear least-squares solver and then add the derived constant offsets to each CCD image.

Next, for each massive galaxy and CCD, we define a circular annulus whose inner and outer radii are a multiple of the galaxy characteristic radius (which is taken as 250 kpc for all galaxies in the intermediate-$z$ sample), and measure the median sky background in that annulus after aggressively and iteratively rejecting the pixels contaminated by astrophysical sources. Finally, we subtract that pedestal sky value as measured above from the whole CCD.

To determine the optimal choice of the sky annulus, we select a sample of 200 intermediate-$z$ galaxies (approximately 10\% of the sample) with well-determined HSC $r$-band surface brightness profiles.  We investigate a range of annuli, 1.0-1.1, 1.1-1.2, 1.2-1.3, 1.3-1.4, and 1.4-1.5 times the cluster radius, and compare the resulting surface brightness profiles to one another and to the corresponding HSC surface brightness profile. Based on these tests, we find that a sky annulus which is 1.0-1.1 times the cluster radius yields the most robust outer surface brightness profiles and galaxy curves of growth.

%To measure the central-galaxy surface brightness profiles we use the ellipse-fitting tools in \href{https://photutils.readthedocs.io/en/stable}{\code{photutils}}.\footnote{\url{https://photutils.readthedocs.io/en/stable}}  First, we project the background-subtracted CCD images onto a common tangent plane and pixel scale ($0\farcs262$ per pixel) using a Lanczos-3 (sinc) interpolation kernel, and build the image stack (in each bandpass) using inverse variance weighting.  Next, we use \tractor{} to model and subtract from the mosaic all the sources in the field \textit{except} the central galaxy of interest (see Section \ref{subsec:2Dmodel}, \citealt{Dey2019},  and J. Moustakas et al., in preparation). In general, the fitting of the surrounding galaxies (inclusing the cluster satellites) and foreground stars in the DECaLS imaging is reliable; nevertheless, to account for model mismatch and other systematic residuals, we mask outlier pixels identified in the residual images (excluding the central galaxy). Finally, we fix the geometry (position angle and ellipticity) of each elliptical isophote to the values derived from the HSC imaging (see Section \ref{sec:SBP-method}) and measure the median surface brightness (iteratively clipping $3\sigma$ outliers) along the ellipse at each semi-major axis position.

Finally, we measure the surface brightness profile of the central galaxy. First, we project the background-subtracted CCD images onto a common tangent plane and pixel scale ($0.262\arcsec$ per pixel) using a Lanczos-3 (\code{sinc}) interpolation kernel, and build the image stack (in each filter) using inverse variance weighting.  Next, we use \tractor{} to model and subtract from the mosaic all the sources in the field \textit{except} the massive galaxy of interest at the center (see Appendix \ref{subsec:2Dmodel}, \citealt{Dey2019},  and J. Moustakas et al., in preparation). In general, the fitting of the surrounding objects (including the satellite galaxies) and foreground stars in the DECaLS imaging is reliable; nevertheless, to account for model mismatch and other systematic residuals, we mask outlier pixels identified in the residual images (excluding the central galaxy). Finally, we fix the geometry (position angle and ellipticity) of each elliptical isophote to the values derived from the HSC profiles (see Section \ref{sec:SBP-method}) and measure the median surface brightness (iteratively clipping $3\sigma$ outliers) along the semi-major axis using the elliptical isophotal analysis tools in \href{https://photutils.readthedocs.io/en/stable}{\code{photutils}}.\footnote{\href{https://photutils.readthedocs.io/en/stable/isophote_faq.html\#how-reliable-are-the-fluxes-computed-by-the-ellipse-algorithm}{Extensive independent tests} show that the surface brightness profile calculated by \code{photutils} agrees with IRAF {\ellipse} result to a very high precision (deviation less than 0.1\%).}

\subsection{Dragonfly Methodology}
\label{sec:dragonfly_methods}

In this section we briefly describe the sky background modeling adopted for the Dragonfly Wide Field Survey \citep{Danieli2020}. We refer the interested reader to \citet{ZhangThesis2018} and \citet{Danieli2020} for more details. 

%The operation of Dragonfly is fully automated. A modularized, fully automated data reduction pipeline was designed and built to produce photometric and astrometric calibrated, flat-fielded coadd images in both $g$- and $r$-band. First, a series of tests are done to filter out bad dark and flat frames, then a master dark frame and a master flat frame are generated based on remaining good frames. Next, bad science frame---such as those having tracking issues, taken under cloudy skies, out of focus or having pointing errors---are excluded. The zeropoint of each good science frame is calculated by comparing point source magnitude with the AAVSO Photometric All-Sky Survey (APASS) catalogue \citep{Henden2016} after extinction correction.

Sky modeling (and subsequent subtraction) is an important step of the Dragonfly data reduction pipeline. The pipeline is designed to conserve as much low surface brightness emission as possible. A two-stage sky subtraction is applied. In the first stage, \code{SExtractor} is used to create a background map for each science exposure with a background mesh size of $128 \times 128$ (\code{BACK\_SIZE=128}). Then the background map is fitted by a third-order polynomial and subtracted from each individual exposure. Next, the images are registered to a common grid with pixel scale of $2.5\arcsec$ using \code{SCAMP} \citep{SCAMP} and \code{SWarp} \citep{Swarp} and scaled to a common flux level. Median $g$ and $r$-band coadd images are generated by taking the median of all sky-subtracted frames in each band. The summed $g+r$ median coadd thus represents the full detection power of the data set. In the second stage, the sky background of a single frame is modeled in the same way but using an input weight map for the masking stage. The weight map is generated from the corresponding area in the $g+r$ deep coadd image, in which sky pixels have the value of 1 and source pixels (to be masked) are 0. Masking individual science frames using the deep coadd ensures that all sources are masked down to their very low surface brightness edges. Then a sky model is fitted and subtracted from each individual frame. Cosmic rays and satellite trails are also removed from each frame. At last, single frames are co-added together in a weighted-average way to optimize the signal-to-noise ratio of the final image, where the weight of each frame is inversely proportional to both the sky background brightness and the scaling of zeropoint. The co-added image is also projected with pixel scale $2.5\arcsec$ per pixel. Under this meticulous sky subtraction, the Dragonfly Wide Field Survey achieves a $1\sigma$ surface brightness detection limit of $\mu_{r} = 29.6\ \mathrm{mag/arcsec}^2$, $\mu_{g} = 29.2\ \mathrm{mag/arcsec}^2$ on scales of $10\arcsec$.

The Dragonfly Telephoto Array has already shown  tremendous potential for detecting extended low surface brightness phenomena \citep{vanDokkum2014, Merritt2016a, Merritt2016b, Zhang2018, Cohen2018, Miller2021,Keim2021}. However, Dragonfly has the worst seeing (FWHM $\sim 5$ arcsec) and lowest resolution ($2.5\arcsec$ per pixel) among the four surveys, which makes blending a major issue in Dragonfly data. For the study of low surface brightness outskirts, the scattered light from nearby stars and galaxies often masquerade as low surface brightness features. Hence, it is crucial to accurately subtract compact objects and their associated scattered light from Dragonfly images. We mitigate these issues using multi-resolution filtering\footnote{\url{https://github.com/AstroJacobLi/mrf}} (MRF, \citealt{MRF}). The algorithm models and subtracts compact objects based on high-resolution images [e.g. Canada France Hawaii Telescope (CFHT), DECaLS, SDSS], conserving target galaxies and low surface brightness features below a certain magnitude level. The algorithm also models halos around bright stars by stacking bright stars. MRF has already shown its potential in Dragonfly data \citep{vanDokkum2019, Gilhuly2019, Danieli2020, Miller2021, Keim2021}. We apply MRF to the Dragonfly images, mask out residual pixels, and extract 1-D surface brightness profiles using the same methodology as in Section \ref{sec:SBP-method}.

\section{Results}\label{sec:result}

In this section, we compare the surface brightness profiles measured from HSC, DECaLS, SDSS and Dragonfly for the low-$z$ sample (Section \ref{subsec:low-z-sample}). We also compare the surface brightness profiles and stellar mass measurements from HSC and DECaLS for intermediate redshift galaxies (Section \ref{subsec:mid-z-sample}). 
In Section \ref{subsec:linear-profile}, we show surface brightness profiles from HSC and DECaLS in linear flux density units to highlight the capabilities of both surveys for measurements of the outskirts of massive galaxies. In Section \ref{subsec:topn}, we use weak gravitational lensing analysis to demonstrate the consistency on measuring outer mass between HSC and DECaLS.

\subsection{Low Redshift Sample}\label{subsec:lowz-result}

We compare the surface brightness profiles measured from SDSS, DECaLS, HSC, and Dragonfly for the five low-$z$ massive galaxies (see Section \ref{subsec:low-z-sample}). The 1-D surface brightness profiles are extracted and corrected for filter differences based on the methodology described in Section \ref{sec:methodology} and Appendix \ref{subsec:filt-corr}. To remove satellite contamination, we use the binary mask method rather than 2-D modeling (see Appendix \ref{subsec:2Dmodel}) since it provides robust profiles and is more efficient.

The surface brightness profiles of low-$z$ galaxies still suffer from over-subtraction in HSC \code{S18A} even the global background residual is positive in the $r$-band. The raw surface brightness profiles all truncate dramatically in the outskirts. This is not surprising given the large angular size of these bright galaxies\footnote{The HSC CCDs are $4096 \times 2048$ pixels in size. At $z=0.02$, 1024 pixels corresponds to about 69 kpc. So the full extent of the light distribution of these low-$z$ galaxies is often close, or larger than, the half-width of the CCD.}. The bright and extended outskirts of these galaxies are still confused with the local background estimate by \texttt{hscPipe} and subtracted away.
Here we measure the sky background locally (see Section \ref{sec:bkg-mock-test} for details) and correct profiles directly, rather than using \skyobjs{} since the sparsely sampled \skyobjs{} are not able to capture the small-scale variation of backgrounds around the galaxy. For other surveys, we also measure the local sky as above and correct the profiles.

\begin{figure}
	\centering
	\includegraphics[width=1\linewidth]{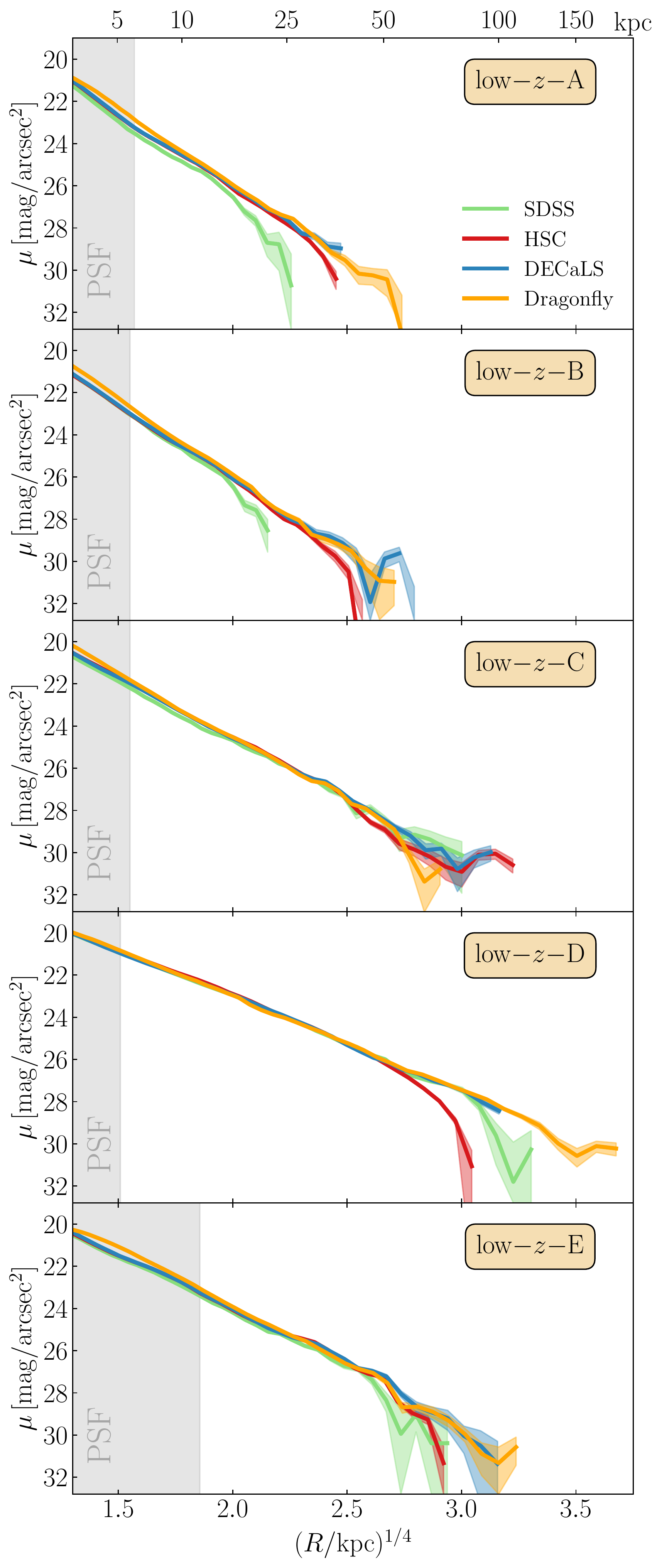}
	\vspace{-20pt}
	\caption{
	$r$-band surface brightness profiles of the low-$z$ sample for four different surveys (thick lines). The profiles are shown using a $R^{1/4}$ scaling on the $x$-axis to emphasize the outskirts. The corresponding $R$ in linear scale is shown on the top axis. The shaded area around each profile indicates the uncertainties given by \ellipse{} (or \code{photutils} for DECaLS). The gray shaded region marks the size of the Dragonfly PSF. The four data sets agree with each other in the inner region. Differences in the outskirts reflect systematic differences in sky subtraction.} %In the outskirts, SDSS profile shows strong over-subtraction signature for all cases. Even with a careful sky background subtraction, HSC pipeline still over-subtracts faint lights in the outermost region for nearby bright galaxies. \jiaxuan{Describe DECaLS profile here.} Among others, Dragonfly provides the most robust and reasonable light profiles, thanks to its optical design and dedicated reduction pipeline
	\label{fig:low-z-main-fig}
\end{figure}

%Small levels of sky over-subtraction (often the case for low-$z$ bright galaxies) can lead to negative surface intensities in the outer regions. This will lead to a sudden break of the profile expressed in $\mathrm{mag/arcsec}^2$. To make the profile more reasonable and convey more information, we define an upper-limit profile by adding the $1\sigma$ standard error of median sky value (i.e. the background uncertainty, as defined in Section \ref{sec:bkg-mock-test}) to the surface brightness profile. A lower-limit profile can also be defined in the same way, but it will be ill-defined when the observed intensity is smaller than the uncertainty of background, in which case we are concerned only about the upper-limit profile. A measured surface brightness profile can be in any place between the upper-limit and lower-limit profiles. \alexie{Song, can you please edit this section to make it more clear?} \jiaxuan{We no longer show this error.}

%In this work, we estimate the sky uncertainty using the standard deviation of the intensity distribution of background super-pixels after masking and binning (see Section \ref{sec:bkg-mock-test}). 

Figure \ref{fig:low-z-main-fig} shows the 1-D surface brightness profiles of the low-$z$ sample. The shaded area around each profile indicates the uncertainties of surface brightness given by \ellipse{} (or \code{photutils} for DECaLS). The vertical grey region corresponds to three times the typical FWHM of the Dragonfly PSF, which is the largest among these data sets. Since the low-$z$ galaxies have moderate stellar mass ($10.5 < \log M_{\star}/M_{\odot} < 11.1$), the surface brightness profiles typically extend to $\sim 50$ kpc at the detection limit. For the more massive galaxies (\code{low-z-D}, \code{low-z-E}), their profiles extend beyond 100 kpc. The surface brightness profiles show good agreement in the inner regions (typically $10\ \mathrm{kpc} < R < 30\ \mathrm{kpc}$), but exhibit large differences in the outskirts (larger than the statistical uncertainties of individual profiles).  

It is well-known that SDSS significantly over-subtracts the background around bright objects (e.g., \citealt{Bernardi2007,Blanton2011}). We also find that the SDSS profiles of low-$z$ galaxies show early artificial truncation in their outskirt (\code{low-z-A}, \code{B}, \code{D}, and \code{E}).
For HSC \code{S18A}, even though we attempt to empirically correct the local over-subtraction for the 1-D profile, we still see over-subtractions in HSC profiles (especially in \code{low-z-D}).
We also tested background correction using \skyobjs{} near low-$z$ galaxies but did not find noticeable improvements. 
Since the over-subtraction happens at the single--CCD level and depends on both the size and brightness of galaxies, it is difficult to mitigate via an empirical sky correction at the coadd level. The correction can only ensure the 1-D profile does not become negative within a certain scale, but cannot fully recover the real outer profile.

%Although HSC uses global sky subtraction to preserve low surface brightness features, we still see over-subtractions in HSC profiles (especially in \code{low-z-D}), even after trying to correcting the over-subtraction by manually estimate the local sky background around the low-$z$ galaxies. This indicates that the over-subtraction in HSC \code{S18A} is hard to mitigate . 

Meanwhile, although DECaLS images are shallower than HSC, the DECaLS profiles of low-$z$ galaxies are as extended as those from HSC and do not truncate at large scales. This is a result of the detailed small-scale background matching and subtraction on CCD level before generating coadd images (see Section \ref{sec:decals_methods}).

At these low redshifts, Dragonfly is the most robust data set against sky subtraction and displays extended surface brightness profiles without evidence of sudden nonphysical truncation. Dragonfly's profiles agree well with DECaLS and HSC profiles to $\sim 29 - 30\, \mathrm{mag/arcsec^2}$, but extend farther than any other surveys. Despite worse seeing and spatial resolution, Dragonfly has an outstanding ability to detect low surface brightness features to a deeper levels than $30\,\mathrm{mag/arcsec^2}$ owing to the well-controlled PSF, globally modeled sky, and dedicated reduction pipeline including the MRF technique.

From this comparison, we find that both DECaLS and Dragonfly can reliably probe the outer regions of massive low-$z$ galaxies, reaching $\sim 29\ \mathrm{mag/arcsec}^2$ in $r$-band without significant over-subtraction. The key challenge with accurate photometry for low-$z$ bright galaxies is to overcome the over-subtraction of the sky background. The comparison between HSC and DECaLS highlights the fact that  
deep exposures alone do not guarantee advantages when studying the low surface brightness regime. One must rectify the sky background at the single-exposure level (see \citealt{Fliri2016}).
%Although HSC is deeper than any other surveys on point-source detection, it does not imply HSC is the best for studying galaxy outskirts, unless the background is carefully modelled and subtracted. 

In practice, small levels of sky over-subtraction can lead to negative surface intensity, which causes a sudden break in the profile when expressed in logarithmic units (such as $\mathrm{mag/arcsec}^2$). In Section \ref{subsec:linear-profile} we discuss the advantage of using a linear scaling to study light profiles, which enables us display information carried by the negative surface intensities. 

\subsection{Intermediate Redshift Sample}\label{subsec:midz-result}

\begin{figure*}
	%\hskip -5mm
	\vbox{ 
		%\vskip -10mm
		\centering
		\includegraphics[width=0.98\linewidth]{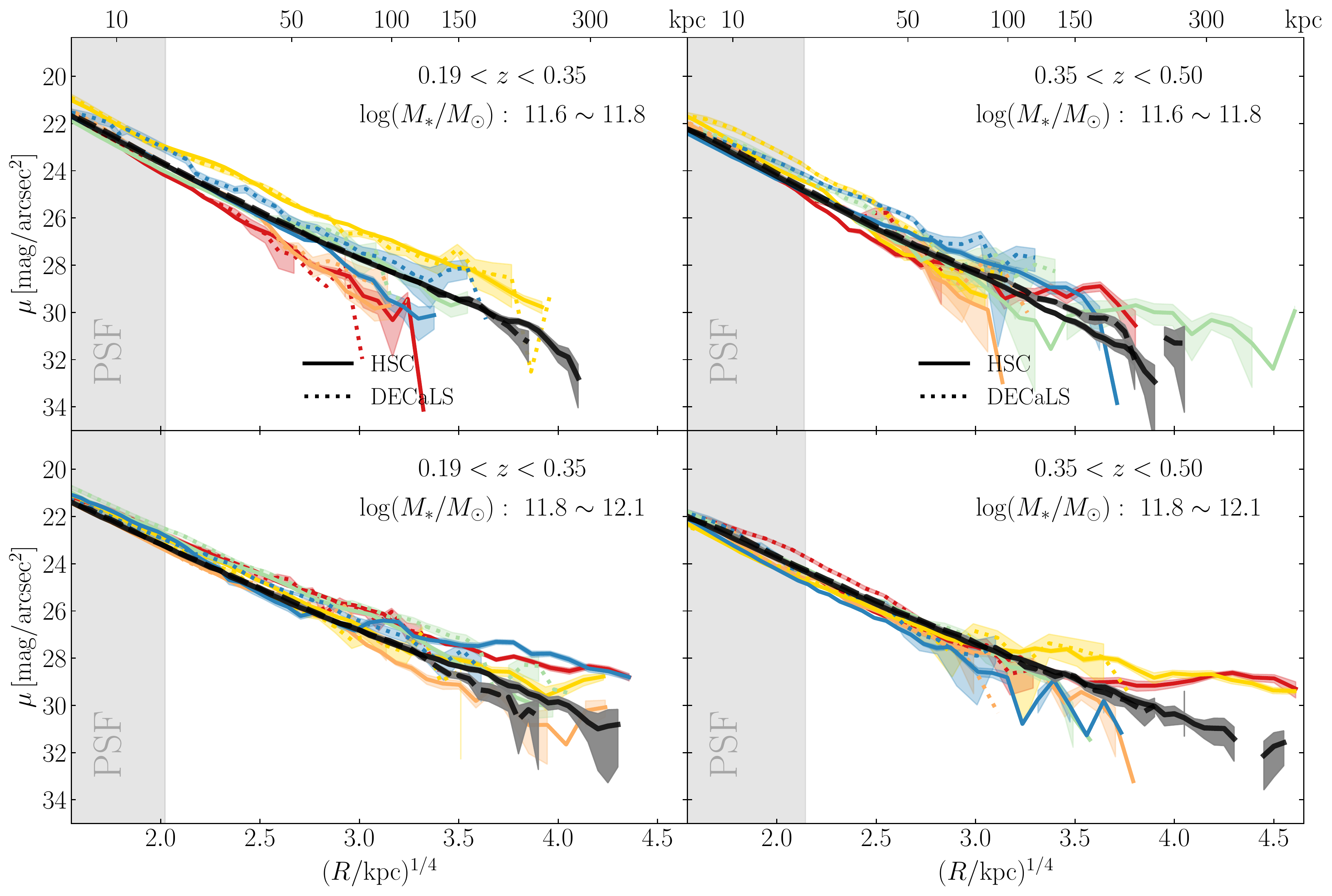}
	}
	\caption{$r$-band surface brightness profiles of intermediate-$z$ massive galaxies for HSC (solid lines) and DECaLS (dotted lines). Five randomly selected profiles are shown with colored lines, together with the median profiles of each mass and redshift bin (thick black lines). HSC and DECaLS median profiles agree with each other within $R \sim 100\ \mathrm{kpc}$. Individual surface brightness profiles have large fluctuations in the outskirts meaning that it is difficult to directly compare individual profiles. 
	}
	\label{fig:mid-z-main}
\end{figure*}

We only use HSC and DECaLS images for the intermediate redshift sample since the redshift range is too high for Dragonfly (because the PSF is too large) and for SDSS (as the images are too shallow). We divide the intermediate-$z$ sample into two redshift bins ($0.19 < z < 0.35$ and $0.35 < z < 0.50$) and two stellar mass bins ($11.6 < \log(M_*/M_\odot) < 11.8$ and $11.8 < \log(M_*/M_\odot) < 12.1$). In the lower redshift bin, there are 711 (125) galaxies in the lower (higher) stellar mass bin. In the higher redshift bin, there are 994 (215) galaxies in the lower (higher) stellar mass bin. The choices of redshift and stellar mass bins do not affect any results in this work.  

HSC and DECaLS surface brightness profiles are extracted according to the methodology described in Section \ref{sec:hsc_methods} and Section \ref{sec:decals_methods}, respectively. For each galaxy, the DECaLS profile shares the same fixed shape (ellipticity and position angle, which are determined from the HSC profile) with the HSC profile. As described in Section \ref{sec:bkg-mock-test}, Section \ref{subsec:linear-profile} and Appendix \ref{appendix:mock-test}, we correct all HSC profiles  with $8.4\arcsec$ aperture {\skyobjs}. We apply filter difference correction (see Appendix \ref{subsec:filt-corr}) to every DECaLS profile.
We summarize the main results in Figure \ref{fig:mid-z-main}.

Figure \ref{fig:mid-z-main} show the surface brightness profiles for the different redshift and mass bins. We use three times the FWHM of the DECaLS seeing (in radius) to indicate the central regions affected by the smearing effect (shaded region).

We show the HSC (solid lines) and DECaLS (dotted lines) profiles of five randomly selected galaxies in each redshift bin and stellar mass bin (colored lines) to demonstrate the typical quality of the surface brightness profile of individual galaxies. 
Since the intermediate-$z$ sample contains mostly very massive galaxies ($\log M_{\star}/M_{\odot} > 11.5$), both the HSC and DECaLS $r$-band images have no problem securing individual 1-D profile to $\sim 100$ kpc. For many galaxies, the 1-D profile can be well measured out to $\sim 150$ kpc.
The HSC and DECaLS individual profiles usually agree with each other reasonably well within $\sim 50$ kpc, but sometimes show large differences at larger radii.

We also display the median profile for each bin (thick black line).
We estimate the statistical uncertainty of the median profile (shaded regions) by bootstrap resampling. The noisy nature of over-subtracted background can cause negative flux density values, which result in undefined surface brightness values (\code{NaN}). When generating median profiles, we do not remove any individual profiles. If more than 80\% of the surface brightness profiles at a given radius $R$ are undefined, we do not show the median profile at $R$ because of too few profiles at this radius. Changing this fraction to 60\% does not alter our results.

%we discard a galaxy when $>30$\% of the data points are undefined \song{30\% sounds a little arbitrary to me since this the number of data points in a profile depends on the sampling parameters and are different between HSC and DECaLS. Can we say if we change this fraction to 20\% or 40\%, nothing will change?}. 
%In total, we remove XX galaxies from HSC and XX galaxies from DECaLS \song{Need to report the number. It should be a small number so we can say it has no effect on the results. We should also check their images. They should be around bright objects like saturated stars.}.

In general, the median profiles of HSC and DECaLS agree well with each other to $\sim100$ kpc ($\sim 28\ \mathrm{mag/arcsec^2}$). 
In the outer region, the median surface brightness profiles show large discrepancies outside of $\sim 100$ kpc in two bins.
The HSC median profile is systematically higher in the $0.35 < z < 0.50$ and $11.6 < \log(M_*/M_\odot) < 11.8$ bin (top right) than DECaLS, while it becomes lower than DECaLS in the $0.19 < z < 0.35$ and $11.8 < \log(M_*/M_\odot) < 12.1$ bin.
Note that, for these intermediate-$z$ galaxies, $> 100$ kpc is often beyond the reach of DECaLS image for any single galaxy and many HSC and DECaLS profiles still have undefined data points at $> 100$ kpc. Although the median profiles are calculated using the same sample, every data point in the median profile is actually based on different sets of galaxies, due to the undefined profile issue. Therefore, we think the differences between median profiles seen in Figure \ref{fig:mid-z-main} is caused by the fact that different sets of galaxies are used to estimate the median profiles.
In fact, this is related to the choice of showing surface brightness profile in magnitude units. We will discuss this point further in Section \ref{subsec:linear-profile}.

Although HSC and DECaLS median profiles are consistent with each other within $\sim 100\ \mathrm{kpc}$ ($\mu_r\sim 28.5\ \mathrm{mag/arcsec^2}$), it is important to remember that the surface brightness profile of any single galaxy is still noisy in the outskirts. One should be very cautious when comparing individual profiles, especially in the low surface brightness regime.

\subsection{Comparisons of Outer Profiles using Linear Scaling}
    \label{subsec:linear-profile}

\begin{figure*}
	%\hskip -5mm
	\vbox{ 
		%\vskip -10mm
		\centering
		\includegraphics[width=1\linewidth]{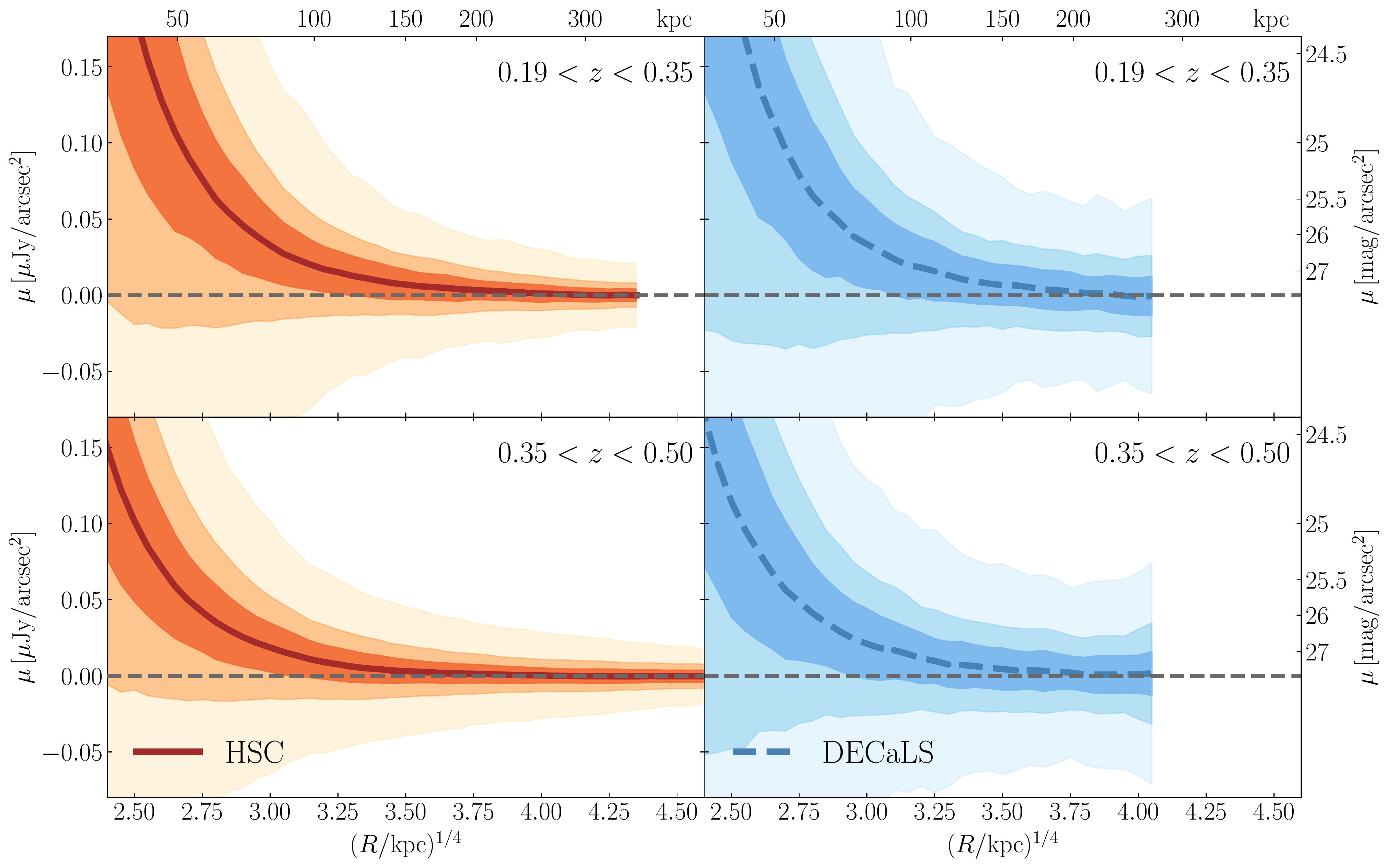}
	}
	\caption{$r$-band surface brightness profiles of the intermediate-$z$ sample displayed using intensity units $\mu\mathrm{Jy/arcsec^2}$ (instead of magnitudes). The linearized profiles better demonstrate the accuracy of the sky background subtraction and highlights the difference between HSC and DECaLS. The median profile is shown in thick lines, and the three shaded regions around the median profile indicate the $0.5\sigma$, $1\sigma$ and $2\sigma$ standard deviation of the profile distribution. After carefully correcting for the sky background, the median profiles of both HSC and DECaLS converge to zero smoothly, indicating that HSC and DECaLS show no systematic bias on sky subtraction. It is also noticeable that the scatter for DECaLS profiles is larger than for HSC.
	}
	\label{fig:mid-z-linear}
\end{figure*}

\begin{figure*}
	%\hskip -5mm
	\vbox{ 
		%\vskip -10mm
		\centering
		\includegraphics[width=1\linewidth]{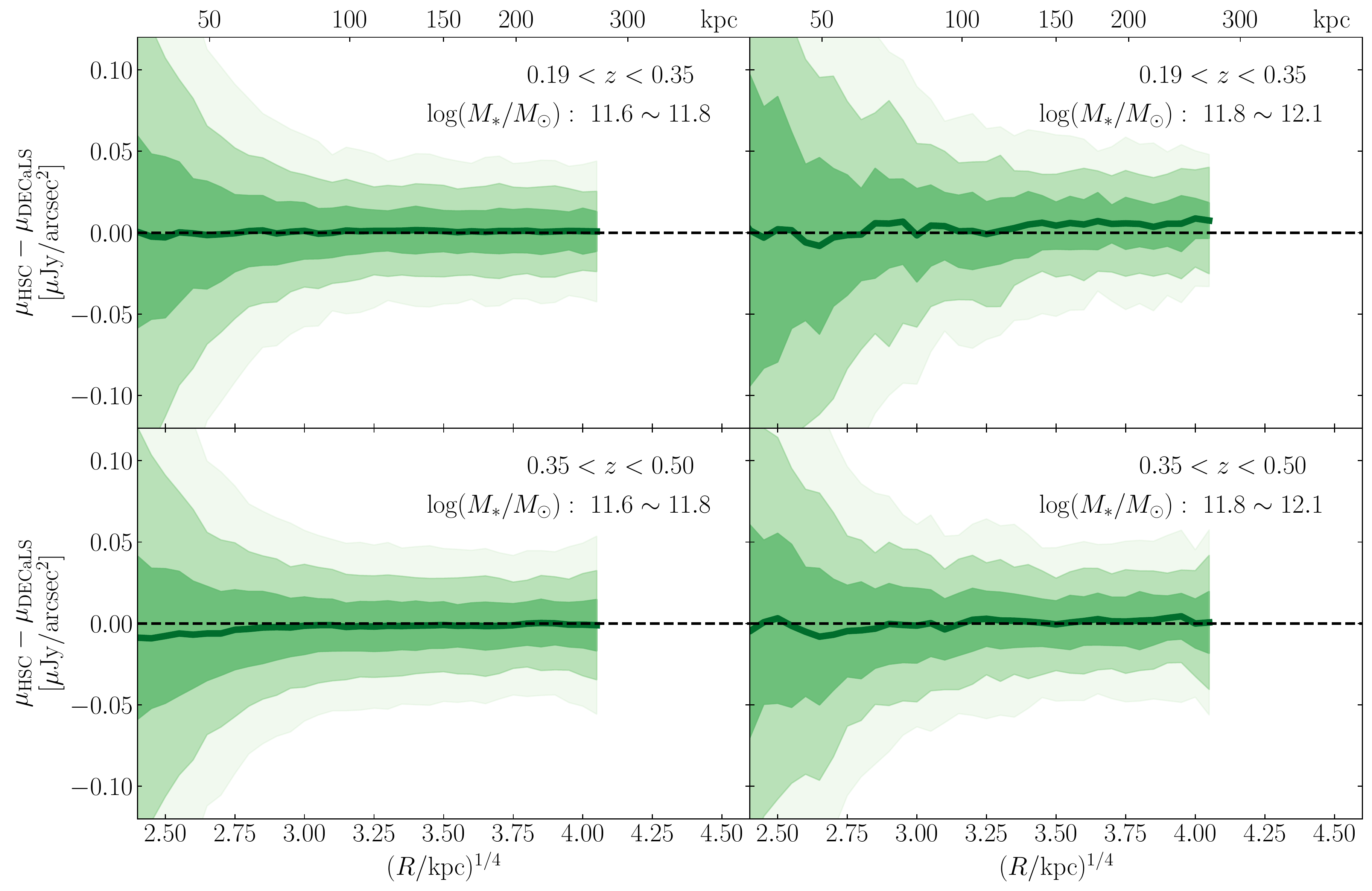}
	}
	\caption{Difference between HSC and DECaLS profiles for	the intermediate-$z$ sample ($\mu_{\mathrm{HSC}}-\mu_{\mathrm{DECaLS}}$ as a function of radius) in linear units $\mu\mathrm{Jy/arcsec^2}$. The median of the difference profile is shown with a thick line, the three shaded regions around the median profile indicate the $0.5\sigma$, $1\sigma$ and $2\sigma$ standard deviation of the distribution. After careful sky subtraction and effective filter correction (Appendix \ref{subsec:filt-corr}), HSC and DECaLS profiles of individual galaxies are consistent. Together with Figure \ref{fig:mid-z-linear}, this demonstrates that HSC and DECaLS have no systematic bias with regards to sky subtraction and are statistically consistent with each other in terms of their light profile measurements. 
	}
	\label{fig:mid-z-linear-diff}
\end{figure*}
In this subsection, we further compare the HSC and DECaLS profiles of the intermediate-$z$ sample with a focus on the outermost region. We also highlight the advantage of using linear flux density units instead of magnitudes.

While over-subtraction and noise can lead to negative flux densities that are undefined in magnitude space, we incur a bias by simply removing them because the negative flux densities contain important information about sky background and noise. In Figure \ref{fig:mid-z-linear}, we show the 1-D intensity profiles of the intermediate-$z$ galaxies in two redshift bins using linear units of $\mu\mathrm{Jy/arcsec^2}$ without removing any galaxy. In this way, for a given $R$, the median profiles in HSC and in DECaLS are calculated based on the same subset of galaxies. 
We only show the $R >30$ kpc profiles to emphasize their behavior in the outermost region. The thick lines in Figure \ref{fig:mid-z-linear} highlight the median profiles. 

The median HSC and DECaLS profiles slowly and smoothly converge to zero between 200 and 300 kpc without any sign of systematic over-subtraction. 
Also, the distribution of intensity profiles becomes symmetric around zero in the outermost region, suggesting that the flux density levels are controlled by the statistical background noise level. 
Both of these results indicate that, on average, there is no systematic bias due to the background subtraction for either HSC or DECaLS. As can be seen in Figure \ref{fig:mid-z-linear}, the shallow nature of the outer intensity profile of massive elliptical galaxies makes the accuracy of background subtraction extremely important when studying the low surface brightness regime. This requires sometimes customized approaches. For example, full focal plane background subtraction plus empirical correction using \skyobjs{} for HSC, and background subtraction at the single exposure level for DECaLS. We also test different \skyobjs{} aperture sizes within HSC, and find that correction with $8.4\arcsec$ aperture \skyobjs{} gives the best convergence to zero for our sample.

For HSC, the median profile for the $0.19 < z < 0.35$ galaxies reaches zero intensity at close to 280 kpc while the one for the $0.35 < z < 0.5$ bin stops at around 230 kpc. Given the similar stellar mass distributions in both bins and the redshift separation is small, this most likely reflects the cosmology dimming effect of surface brightness. At $R=300$ kpc, the scatters of intensity are 0.009 $\mu\mathrm{Jy/arcsec^2}$ (0.011 $\mu\mathrm{Jy/arcsec^2}$) for $0.19 < z < 0.35$ ($0.35 < z < 0.5$) bin, broadly consistent with the tests using \skyobjs{} (where $\sigma_r = 0.008\ \mu\mathrm{Jy/arcsec^2}$, see Appendix \ref{fig:skyobj_stats}). 
For DECaLS, the median profiles for both bins approach zero intensity at around 200 kpc and show larger scatter values (0.039 $\mu\mathrm{Jy/arcsec^2}$ and 0.035 $\mu\mathrm{Jy/arcsec^2}$ for each redshift bin), consistent with what we expect from its shallower images.

We further compare HSC and DECaLS profiles on a galaxy-to-galaxy basis using a linear intensity scaling.
In Figure \ref{fig:mid-z-linear-diff}, we show the differences between individual intensity profiles for the intermediate-$z$ galaxies in all four redshift and stellar mass bins.
In all bins, the differences profiles (i.e., $\mu_{\mathrm{HSC}}-\mu_{\mathrm{DECaLS}}$ in $\mu\mathrm{Jy/arcsec^2}$) are statistically consistent with zero within the 1$\sigma$ uncertainties. The uncertainty of $\mu_{\mathrm{HSC}}-\mu_{\mathrm{DECaLS}}$ (0.03 $\mu\mathrm{Jy/arcsec^2}$ at 200 kpc) also roughly agrees with the background uncertainty of DECaLS profiles. 

This result indicates that the HSC and DECaLS profiles of intermediate-$z$ massive galaxies statistically agree with each other in the low surface brightness regime once the background subtraction has been carefully calibrated. 
This implies that one can study the outskirts of $z<0.5$ galaxies using DECaLS images with the same statistical accuracy as  the deeper HSC images while enjoying the much larger sample size that the DECaLS offers due to the wide area of the DECaLS survey. 

Finally, we advocate for the usage of linear flux density units such as $\mu\mathrm{Jy/arcsec^2}$ for profiles in the low surface brightness regime or close to the image detection limit.
The linear flux density can naturally extend to negative values that contain useful information regrading the sky background and image noise level. It overcomes the issue of undefined surface brightness in magnitude (logarithmic) units. Moreover, in the low surface brightness regime, a tiny shift of flux density in $\mu\mathrm{Jy/arcsec^2}$ can lead to dramatic changes in mag/arcsec$^2$ units. For example, at $\mu_{\rm HSC} = 28.0$ mag/arcsec$^2$, an increase (decrease) of 0.005 $\mu\mathrm{Jy/arcsec^2}$ in flux density will change the surface brightness to 0.7 (1.7) mag/arcsec$^2$. Linear flux density unit provides a more robust representation of surface brightness profiles.

\subsection{Robustness of Mass Estimates}
    \label{subsec:stellar-mass}

\begin{figure*}
	%\hskip -5mm
	\vbox{ 
		%\vskip -10mm
		\centering
		\includegraphics[width=0.95\linewidth]{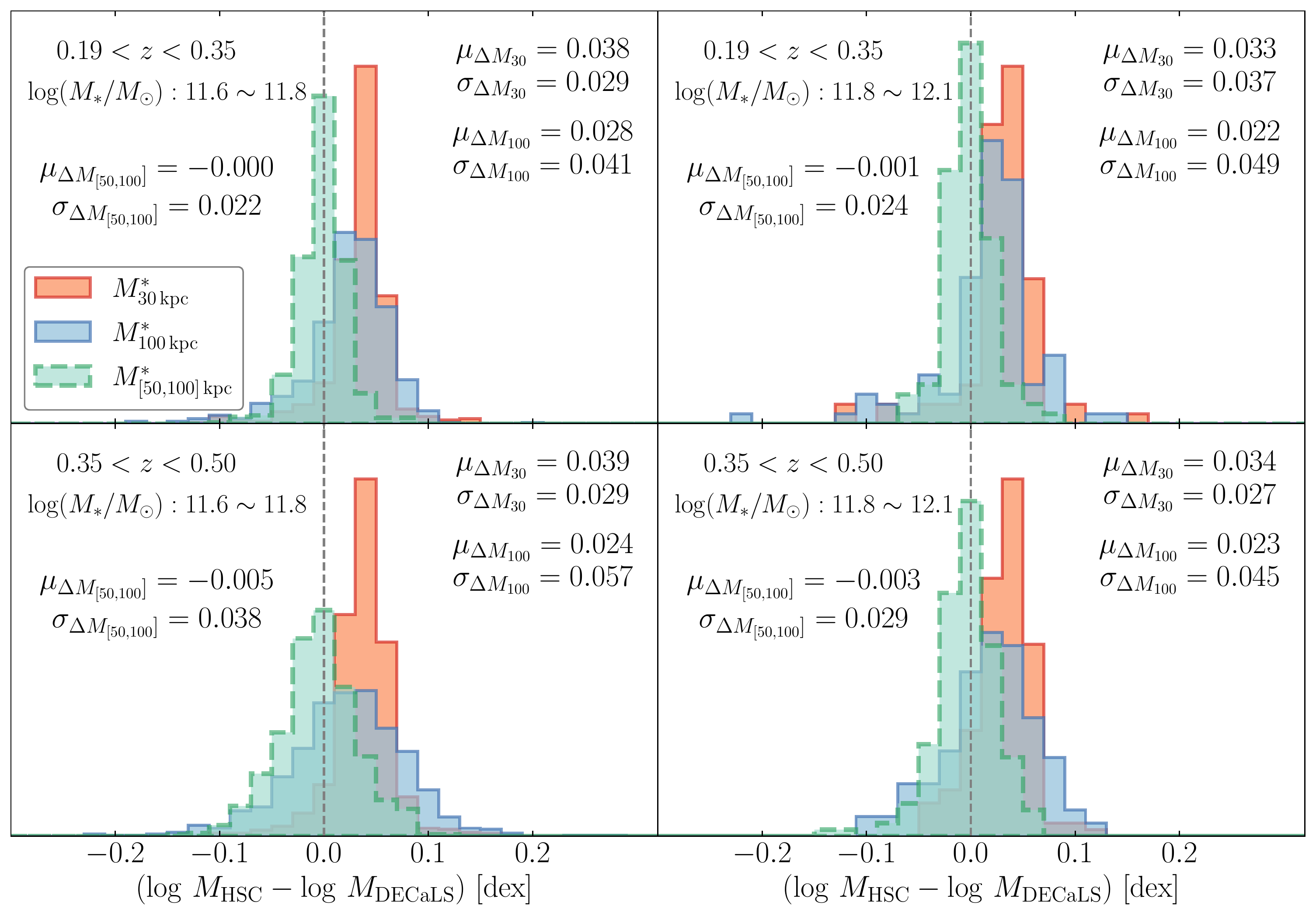}
	}
	\caption{Histograms of stellar mass differences ($\Delta M = \log M_{\mathrm{HSC}} - \log M_{\mathrm{DECaLS}}$) measured by HSC and DECaLS. We compare the stellar mass within 30 kpc ($M^*_{30\,\mathrm{kpc}}$) and 100 kpc ($M^*_{100\,\mathrm{kpc}}$) in this figure, as well as the stellar mass between 50 kpc and 100 kpc ($M^*_{[50, 100]\,\mathrm{kpc}}$) as an indicator of the ``outer mass''. Statistically, the stellar masses estimated by the two surveys agree well with each other with a median deviation less than 0.05 dex. The scatter of stellar mass difference is also small (less than 0.05 dex). The distributions of outer mass difference (which is not affected by filter correction between the two surveys) are almost centered at zero, indicating that the mass measurements using HSC and DECaLS are consistent with each other.}
	\label{fig:mass-comparison}
\end{figure*}

One of the main reasons for extracting a surface brightness profile is to estimate the luminosity and stellar mass of a galaxy in a model-independent manner (e.g., \citealt{Huang2018c,Miller2021}). 
In massive galaxies, stellar masses within different apertures (e.g. $M^*_{10\,\mathrm{kpc}}$, $M^*_{30\,\mathrm{kpc}}$, or $M^*_{100\,\mathrm{kpc}}$) could represent different aspects of their mass assembly history (e.g. \citealt{Huang2020ASAP}).
While the inner regions of massive galaxies could mainly consist of stars formed within the halo of the main progenitor (so-called \emph{in-situ} stars), their outskirts are dominated by stars accreted onto galaxies at different times (the \emph{ex-situ} component; see e.g., \citealt{vanDokkum2008,Rodriguez-Gomez2016}).
Recently, \citet{Huang2021} found that the outer stellar mass (defined as the stellar mass between 50 kpc and 100 kpc) is a promising proxy of the dark matter halo mass. Therefore, in this section, we compare the stellar mass measurements within different apertures between HSC and DECaLS for the intermediate-$z$ sample. 

First, we integrate the 1-D HSC and DECaLS profiles, after accounting for the isophotal shape, to estimate the apparent magnitudes within different physical radii $R$.
We correct DECaLS magnitudes to the HSC $r$-band following Appendix \ref{subsec:filt-corr}.
For the same galaxy, the magnitude difference between the HSC and DECaLS images can be directly translated into the luminosity difference. We assume that the mass-to-light ratio ($M/L$) estimated using HSC five-band SEDs can be applied to DECaLS and we infer the difference in $\log M_{\star}/M_{\odot}$ directly from the magnitude difference.

Figure \ref{fig:mass-comparison} compares the HSC and DECaLS values of $M^*_{30\,\mathrm{kpc}}$ (red), $M^*_{100\,\mathrm{kpc}}$ (blue), as well as the outer mass $M^*_{[50,100]\,\mathrm{kpc}} = M^*_{100\,\mathrm{kpc}} - M^*_{50\,\mathrm{kpc}}$ (green). We highlight the median difference in stellar mass (denoted as $\mu$) and its standard deviation ($\sigma$) in Figure \ref{fig:mass-comparison}. 

For $M^*_{30\,\mathrm{kpc}}$ ($M^*_{100\,\mathrm{kpc}}$), we find that the HSC estimates are on average slightly higher than the DECaLS values by 0.03-0.04 dex ($\sim 0.02$ dex) with a small scatter around 0.02-0.04 dex (0.04-0.06 dex).
Both the offset and scatter values are smaller than the typical statistical uncertainties in the HSC stellar mass estimates 
($\sim 0.1$ dex; see \citealt{Huang2018c}).
For $M^*_{[50,100]\,\mathrm{kpc}}$, we instead find that the HSC and DECaLS estimates are on average fully consistent with each other with a small scatter of 0.02-0.04 dex.
Statistically speaking, we find that the aperture or outskirt stellar masses from HSC and DECaLS show no significant difference.

It is interesting to note that the difference stems mainly from the inner rather than the outer regions.
This suggests that it is not the imaging depth that drives the higher stellar mass values for HSC, which is consistent with the results in \S \ref{subsec:midz-result}.
There are several possibilities for this phenomena. 
Firstly, we did not take the PSF smearing effects into account when measuring the aperture stellar mass. The worse seeing of DECaLS images scatters more light toward outer region than the HSC one. Indeed, the stellar mass differences of $M^*_{10\,\mathrm{kpc}}$ and $M^*_{20\,\mathrm{kpc}}$ are larger than that of $M^*_{30\,\mathrm{kpc}}$, indicating the PSF smearing effects do play an important role. 
Also, we derive the filter correction from a small number of stellar spectra and ignore possible color gradients within the massive galaxies.
Imperfect filter correction values could also lead to small offsets when comparing aperture stellar masses.
%\song{What's the scatter of the residual from filter correction? Maybe we can say it is larger than the offset of mass difference?}\jiaxuan{The scatter of filter correction term is 0.02 mag, which is 0.01 dex in stellar mass. Larger than $\sigma_{\Delta M}$.}
However, the filter correction term cancels out when deriving the outskirt stellar mass $M^*_{[50,100],\,\mathrm{DECaLS}}$.
Therefore the comparison of $M^*_{[50,100]}$ between HSC and DECaLS should not be affected too much by filter correction and PSF smearing effects.

A more thorough comparison of stellar mass estimates requires a better treatment of seeing effect, filter correction, color gradients, and inclination-dependent extinction, which are beyond the scope of this paper.
Here, we have demonstrated that the DECaLS $r$-band images can provide statistically consistent aperture and outskirt stellar mass estimates with the deeper HSC ones when using the 1-D profile method. 
We are particularly encouraged by the consistency of the outskirt stellar mass estimates such as $M^*_{[50,100]}$, which are promising proxies of halo mass (\citealt{Huang2021}).
This means we can take advantage of the much larger footprint of DECaLS together with DESI redshifts to further the study of galaxy-halo connection for massive galaxies.

\begin{figure}
	\centering
	\includegraphics[width=\columnwidth]{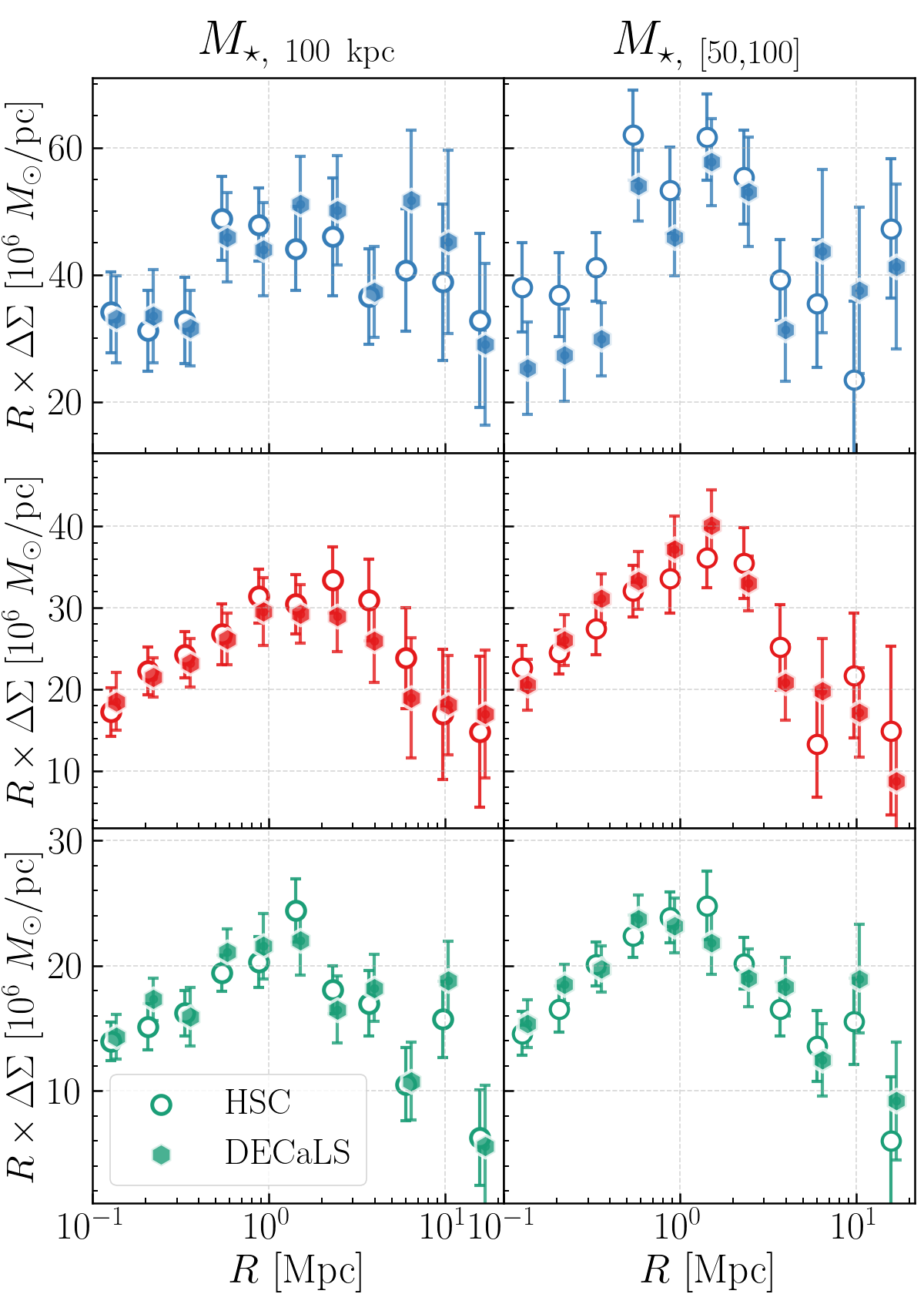}
	\caption{
	    Comparisons of the stacked galaxy--galaxy lensing profiles (using $R \times \Delta\Sigma$) of massive galaxies selected by 
	    rank-ordering different stellar mass estimates based on HSC (empty symbols) and DECaLS (solid symbols) photometry.
	    Left panels show the results for stellar mass within a 100 kpc aperture ($M^{\star}_{100\ \rm kpc}$), while the right panels 
	    are for stellar mass between the 50 and 100 kpc apertures ($M^{\star}_{[50, 100]}$).
	    From top to bottom, the three rows are for the top 50, 50--200, 200--600 most massive galaxies using the corresponding stellar
	    mass estimates.
	    We shift the lensing profiles for the DECaLS galaxies by 7\% in radius for visual purposes. This figure suggests that the outer light method will be a powerful technique for identifying galaxy clusters with DECaLS imaging and DESI redshifts.
	}
	\label{fig:topn}
\end{figure}

\subsection{Top-N test with DECaLS based Measurement of Outer Light}
    \label{subsec:topn}

    \citet{Huang2021} recently showed that the outer stellar mass (e.g., the stellar mass between the 50 and 100 kpc elliptical apertures, $M^*_{[50,100]}$) of massive galaxies is a promising proxy of halo mass with 
    performance comparable to state-of-the-art richness based techniques. \citet{Huang2021} also showed that the outer light method may suffer less from projection effects. 
    However, \citet{Huang2021} use high quality deep HSC data to measure the outer stellar mass. 
    It is of interest to understand whether or not the outer stellar mass measurements from DECaLS are of high enough quality to be used as a halo mass proxy. 
    The most direct way of testing this is using weak gravitational lensing. 
    We follow the exact same methodology as presented in \citet[][]{Huang2021} to perform the so-called ``top-$N$'' test to evaluate the
    performance of DECaLS outer stellar mass estimates. 
    We refer to \citet[][]{Huang2021} for the details of the top-$N$ test.
    In short, we compare the stacked galaxy--galaxy lensing profiles (or the $\Delta\Sigma$ profiles) of samples of massive galaxies selected to 
    share the same number density thresholds based on different stellar mass estimates.
    The overall amplitudes of the $\Delta\Sigma$ profiles reflect the mean value and the scatter of the halo mass of the selected galaxies.
    In practice, we rank-order the intermediate-$z$ sample by different estimates of stellar mass and calculate the stacked $\Delta\Sigma$ 
    profiles of the top 50, 50--200, and 200--600 massive galaxies for each of them\footnote{The definitions of the top-$N$ bins here
    are similar to the Bin 1--3 used in \citet{Huang2021}. However, since the footprints of HSC and DECaLS do not fully overlap, they 
    translate into different number density and stellar mass thresholds than in \citet{Huang2021}}.
    We calculate the lensing profiles using the HSC \texttt{S16A} weak lensing shape catalog \citep{Mandelbaum2018PASJ, 
    Mandelbaum2018MNRAS} and the same methodology as described in Section 4.2 of \citet[][]{Huang2021}. 
    Additional details on how these measurements are performed can also be found in \citet{Speagle2019}, \citet{Huang2020ASAP}, and \citet{Leauthaud2017}.
    
    In Figure \ref{fig:topn}, we compare the lensing profiles of massive galaxies selected based on HSC (empty dots) and DECaLS (solid dots) 
    versions of $M^*_{100\ \rm kpc}$ (left) and $M^*_{[50,100]}$ (right) in the three top-$N$ bins.
    Figure \ref{fig:topn} displays $R \times \Delta\Sigma$ profiles to highlight not only the general amplitudes of the profiles, but also the detailed 
    slopes of profiles in different radial ranges. 
   Figure \ref{fig:topn} shows that, within the uncertainties estimated using Jackknife resampling method, the HSC and DECaLS lensing profiles are
    reasonably consistent with each other. 
    The only notable difference is for the top 50 most massive galaxies using $M_{*, [50,100]}$ (top right panel), where the sample based 
    on HSC photometry shows higher amplitudes inside 1 Mpc. 
    These results suggest that, for the intermediate-$z$ sample tested here, the outer stellar mass estimated based on the shallower DECaLS images has the same statistical capability as a halo mass proxy when compared to HSC\footnote{Note that the intermediate-$z$ sample used here
    is pre-defined using the HSC survey. If we try to independently select massive galaxies based on DECaLS alone in the same footprint, we might end up with different top-$N$ samples that will affect their lensing profiles. We will test this approach in future work.}. This suggests that this outer light method will be a powerful technique for identifying galaxy clusters with DECaLS imaging and DESI redshifts. 
    %\song{Still think we need to mention this, but maybe not here, or just as a footnote?} 

%%%%%%%%%%%%%%%%%%%%%%%%%%%%%%%%%%%%%%%%%%%%%%%%%%%%%%%%%%%%%%%%%%%%%%%%%%%%%%
%     Discussion
%%%%%%%%%%%%%%%%%%%%%%%%%%%%%%%%%%%%%%%%%%%%%%%%%%%%%%%%%%%%%%%%%%%%%%%%%%%%%%
\section{Discussion}
    \label{sec:discussion}

Here we briefly discuss several systematic issues concerning the use of deep imaging surveys to study the low surface brightness outskirts of galaxies along with possible future research directions.

\subsection{Systematic Issues that Affect Measurements of Galaxy Outskirts}
    \label{subsec:concern}
    
\begin{itemize}
    \item \textbf{Sky background uncertainty}: Background estimation remains the most important systematic effects when mapping the outskirts of galaxies. We show that the HSC \texttt{S18A} data can recover the \textit{average} profiles of massive galaxies at $0.2<z<0.5$ out to $>150$ kpc thanks to the global sky correction adopted by the HSC pipeline \citep{HSC-PDR2}. However, the HSC profile of an individual massive galaxy can still be affected by over- or under-subtracted local background at $R > 100$ kpc. The data reduction pipelines of modern imaging surveys are usually not optimized to preserve the extended structure around bright objects. In HSC \code{S18A}, although the global sky subtraction algorithm preserves the faint envelopes of bright objects, it unfortunately makes it incredibly difficult to deblend these areas and produce reliable catalogs for other science. In more recent data releases, including the Public Data Release 3 (\code{PDR3}, \citealt{HSC-PDR3}), the default coadd images and photometric catalogs are based on local sky subtraction. \footnote{But the coadd images with global background subtraction are still available in the data releases.} A carefully designed survey strategy and reduction pipeline (e.g., the Dragonfly survey) or specially tailored pipeline to improve the background subtraction (e.g., the DECaLS survey) is often required to mitigate this issue. As shown in this work, we also highly recommend mock tests and cross-survey comparisons before reaching conclusions regarding galaxy outskirts.
    
    \item \textbf{Low surface brightness structure of the PSF}: PSFs from modern imaging surveys often display a complex outer structure due detector effects (e.g., \citealt{Michard2002, Wu2005, Sirianni1998}), the structure of the telescope or the optical system (e.g. \citealt{Racine1996, Bernstein2007, Slater2009}), and even the atmosphere (e.g. \citealt{DeVore2013}). An ill-characterized PSF `wing' can lead to  wrong or biased conclusions for LSB science \citep{Tal2011, Sandin2014}. Because the usage of red-sensitive thick-CCDs, we often assume the PSFs from HSC, DECaLS and LSST should show a much less prominent outer structure than the PSF of a fully depleted thin-CCD (e.g. SDSS). For HSC images, PSF wings are not a concern when studying the profile of low-$z$ massive galaxies at $r < 29\ \mathrm{mag/arcsec^2}$ (e.g., \citealt{Wang2019}). However, the outer PSF structures in HSC and DECaLS are still not carefully characterized and could hinder us from taking further advantage of the increasingly deep imaging surveys. We will investigate this in future work by stacking non-saturated stars and via comparison to images with a much better controlled outer PSF wing (e.g. Dragonfly, \citealt{Abraham2014,Liu2021PSF}).
    
    \item \textbf{Undetected objects}: Before extracting the 1-D profiles, we rely on aggressive object masks (e.g., the method use here for HSC images) or simultaneously modeling of all detected objects (e.g., the method used here for DECaLS) to reduce the impact from neighboring objects. The two methods show statistical consistency in mock tests (Figure \ref{fig:tractor_mock_test}) and in real galaxies (Figure \ref{fig:mid-z-linear-diff}). 
    In principle, objects that are just below the detection limits can leave their imprint on the surface brightness profiles and potentially bias the intensity level toward higher values in the LSB regime.
    We use sigma-clipping along the isophote to remove pixels with abnormally high/low intensity and also use the median intensity to achieve less biased estimate. As shown in Figure \ref{fig:mid-z-linear}, both HSC and DECaLS median profiles converges to zero in flux density without significant deviation, indicating that the undetected sources did not introduce significant bias in this work. 
    While sigma-clipping and taking the median profile are reasonable approaches for profiles within $100$ kpc (\citealt{Ardila2021}), there is no guarantee that it still works for science in a lower surface brightness regime, where even a tiny bias in mean intensity can impact the results.
    This is especially important for surveys with poorer seeing and/or spatial resolution where faint extended objects can become undetected (e.g., Dragonfly). Multi-resolution filtering \citep{MRF} which utilizes higher resolution survey data is required under this circumstance.
    
    \item \textbf{Galactic cirrus}: It has recently become clear that the diffuse Milky Way dust emission known as the Galactic cirrus is a major challenge when exploring the low surface brightness Universe \citep{Duc2015, Roman2019}. At $\mu_{V} > 26\sim27\ \mathrm{mag/arcsec^2}$, Galactic cirrus start to contaminate extragalactic observations even at high Galactic latitude $b > 75\degree$ \citep{Mihos2017} and could cover up to 80\% of the whole sky down to $\mu_{V}=30\ \mathrm{mag/arcsec^2}$. In light of the next-generation deep imaging surveys (e.g. LSST) and projects designed for LSB studies (e.g., Dragonfly, MESSIER surveyor \citep{MESSIER2017}, Australian Space Eye \citep{ASE2016}, and Huntsman Telescope \citep{Huntsman2019}, it is a pressing task to develop the capability of disentangling Galactic cirrus from low surface brightness extragalactic structures.
\end{itemize}

\subsection{Future Directions}
    \label{subsec:future}
   
    \begin{itemize}
        \item \textbf{Exploring the intra-cluster light}:
            In this work, we describe the outer profiles of massive early-type galaxies irrespective of their dark matter halo mass. However, massive central galaxies of galaxy clusters often host an extremely extended LSB envelope known as the intra-cluster light  (ICL, e.g., \citealt{Carlberg1997, Lin2004, Gonzalez2005, Mihos2005, Mihos2016}). The ICL consists of stars that follow the gravitational potential of the dark matter halo of the host cluster and could represent a crucial tracer of the assembly histories of massive halo (e.g., \citealt{Rudick2006, Rudick2011, Cui2014, Cooper2015}). Using stacked 1-D profiles from the Dark Energy Survey (DES), \citet{Zhang2019} recently revealed a large amount of ICL in the very outskirts of low-$z$ brightness cluster galaxies (BCGs). If true, this result means the stellar mass around these BCGs can double itself outside the radii that can be probed by HSC ($>150$ kpc).
            It could have profound implications for the stellar-halo connection of massive galaxies. We will examine this result by comparing the 1-D profiles of common BCGs in HSC and DES, and by performing careful image stacking analysis using HSC BCG sample. DECaLS shares the same camera with DES and can also provide an independent check using a large sample of nearby clusters (J. Moustakas et al., in preparation).
        
        \item \textbf{Extending to multi-filters}:
            In this work, we use only single-band images to explore the outskirts of massive galaxies. Meanwhile, rest-frame optical color of the stellar halo contains important information about the underlying stellar population \citep{Carollo1993, LaBarbera2012, DSouza2014}.
            Also, $M/L$ profiles derived from multi-band color gradients can help provide more accurate stellar mass density profiles and outskirt stellar mass estimates. However, due to the differences in background subtraction and seeing between two bands, it is extremely difficult to accurately recover the color profiles in the LSB outskirts.
            It requires careful consideration of most of the systematics mentioned in the previous section (e.g., \citealt{Wang2019}). We will first evaluate the reliability of the \emph{average}
            color profile recovered by HSC using mock galaxies.
        
        \item \textbf{The quest for better `total' stellar mass estimates}:
            Accurate and precise stellar mass measurements for massive galaxies are essential for building galaxy-halo connection models. Here we show that the stellar mass of massive galaxies measured within 100 kpc using DECaLS images is consistent with the stellar mass measurements using deeper HSC images at the $\sim 0.05$ dex level (when a customized reduction pipeline is adopted). This careful treatment will also benefit the DESI survey, which will include many massive galaxies in the Luminous Red Galaxy (LRB) and low-$z$ Bright Galaxy Survey (BGS) samples. However, an open question is whether or not the stellar mass distribution of massive galaxies extends beyond 100 or 150 kpc, beyond the reach of the 1-D profile approach. In the quest for an even more accurate `total' stellar mass for these massive galaxies, we will rely on stacked 1-D profiles using HSC and hydro-simulations to explore a reliable way to extrapolate beyond 150 kpc (Leidig et al., in preparation).
            We will also test the potential of 2-D forward modeling to overcome some of the difficulties faced by the 1-D approach (e.g. seeing effect, cannot deal with mergers) and include many systematics issues in the model. We hope that a combination of these methods will help us reach the edges of massive galaxies and to paint a complete picture of their stellar mass distributions.
    \end{itemize}

%%%%%%%%%%%%%%%%%%%%%%%%%%%%%%%%%%%%%%%%%%%%%%%%%%%%%%%%%%%%%%%%%%%%%%%%%%%%%%
%     CONCLUSIONS
%%%%%%%%%%%%%%%%%%%%%%%%%%%%%%%%%%%%%%%%%%%%%%%%%%%%%%%%%%%%%%%%%%%%%%%%%%%%%%
\section{Summary and Conclusions}
    \label{sec:conclusions}

In this paper, we have explored the ability of different data sets to study the outskirts (or the stellar halos) of $z<0.5$ massive galaxies. 
Using images from HSC, DECaLS, SDSS, and Dragonfly, we highlight important effects to consider when extracting 1-D surface brightness profiles in the LSB regime.
We have also performed a series of tests on the robustness of such profiles from the HSC survey. 
We then statistically compare the 1-D profiles of a large sample of $0.19 < z < 0.5$ massive galaxies using both HSC and DECaLS images. Our main results are summarized below.

\begin{itemize}
    \item By comparing HSC, DECaLS, SDSS, and Dragonfly, we find that the main systematic error impacting outer surface brightness profiles is sky subtraction. For HSC, we find that although \code{S18A} (\code{PDR2}) uses a global background calibration approach, filter-dependent sky background residuals are still present in the data (\S \ref{sec:bkg-mock-test}, Fig. \ref{fig:skyobj_stats}, \ref{fig:skyobj-analysis}). We successfully applied a empirical correction for this residual using \skyobjs{} (Fig. \ref{fig:mock-test-compare}). When a  
customized sky subtraction at the CCD level is adopted (\S \ref{sec:decals_methods}), DECaLS shows no significant over- or under-subtraction issues. We also show that mock tests and cross-survey comparisons are useful for understanding the robustness of outer light measures.
    
    \item At low redshift (e.g., $z<0.05$, Fig. \ref{fig:low-z-postage}), Dragonfly has a better performance than other surveys for studying galaxy outskirts. Thanks to its excellent scattered-light control and dedicated pipeline, it can robustly reach down to $\mu_r > 30\ \mathrm{mag/arcsec^2}$ (Fig. \ref{fig:low-z-main-fig}). With a customized sky subtraction, DECaLS can also reliably probe the outskirts to $\mu_r \sim 29\ \mathrm{mag/arcsec^2}$. However, over-subtraction is an issue for HSC and SDSS at $z<0.05$.
    
    \item At intermediate redshift (e.g., $0.19 < z < 0.5$, Fig. \ref{fig:mid-z-distribution}, Fig. \ref{fig:mid-z-postage}), we can reliably extract the 1-D surface brightness profiles of individual massive galaxies down to $\mu_r \approx 28.5\ \mathrm{mag/arcsec^2}$ using HSC \code{S18A} (Fig. \ref{fig:mid-z-main}). DECaLS profiles are consistent with those from HSC but noisier (Fig. \ref{fig:mid-z-linear-diff}). The median surface brightness profiles of HSC (DECaLS) reach to 300 kpc (200 kpc) with no over- or under-subtraction issue (Fig. \ref{fig:mid-z-linear}). 
    
    \item We advocate using linear scaling to study surface brightness profiles, especially at low surface brightness levels (see Fig. \ref{fig:mid-z-linear}, Fig. \ref{fig:mid-z-linear-diff}). The use of magnitudes (or logarithmic scaling) can be misleading and problematic when the measured light profile has negative values (due to the nature of noise). Performing statistics of light profiles using linear scaling (such as $\mu\mathrm{Jy}/\mathrm{arcsec}^{2}$) not only avoids the issue that arises when taking logarithm of negative values in the profiles, but also includes all the information contained in profiles.
    
    \item We compare stellar mass measurements from HSC and DECaLS. We find that HSC and DECaLS are consistent (within 2-sigma) regarding estimates os $M^*_{30\,\mathrm{kpc}}$, $M^*_{100\,\mathrm{kpc}}$, although HSC estimates are on average slightly higher than DECaLS by $0.03$ dex. For the outer mass $M^*_{[50,100]\,\mathrm{kpc}}$, HSC and DECaLS estimates are fully consistent with each other (Fig. \ref{fig:mass-comparison}). We conclude that the outskirts stellar masses measured in HSC and DECaLS show no significant difference. 
    
    \item We further demonstrate the consistency between HSC and DECaLS by comparing the weak lensing signals of massive galaxies selected by mass and at fixed number density. We find that HSC and DECaLS lensing profiles are consistent with each other in terms of their amplitudes (Fig. \ref{fig:topn}). We conclude that outer mass measurement based on DECaLS provide a reliable proxy of halo mass that is comparable to those from HSC \citep[][]{Huang2021} The outer light method will be a powerful technique for identifying galaxy clusters with DECaLS imaging and DESI redshifts.
    
\end{itemize}

Paper II (K. Leidig et al., in preparation) in this series will explore techniques for robustly measuring the light profiles of massive galaxies above 100 kpc. Paper III (J. Li et al., in preparation) will explore image stacking techniques to perform measurements out to larger radii and will delve more deeply into the effect of PSF wings.

% A box errors that I fixed by putting this in temporarily
\section*{Acknowledgments}
% Persons
We thank Pieter van Dokkum and Andy Goulding for insightful discussions. We are grateful for Dustin Lang's efforts in developing DECaLS imaging software, which makes the analysis in this paper possible.

% CSC
The research reported in this paper is supported by the scholarship from China Scholarship Council (CSC) under Grant CSC \# 201806010530 during JL's stay at University of California, Santa Cruz. This material is based on work supported by the U.S Department of Energy, Office of Science, Office of High Energy Physics under Award Number DE-SC0019301. This material is based upon work supported by the National Science Foundation under Grant No.
1714610. AL acknowledges support from the David and Lucile Packard foundation, and from the Alfred P. Sloan foundation. JM gratefully acknowledges support from the U.S. Department of Energy, Office of Science, Office of High Energy Physics under Award Number DE-SC0020086 and from the National Science Foundation under grant AST-1616414. SD is supported by NASA through Hubble Fellowship grant HST-HF2-51454.001-A awarded by the Space Telescope Science Institute, which is operated by the Association of Universities for Research in Astronomy, Incorporated, under NASA contract NAS5-26555.

% HSC part
The Hyper Suprime-Cam (HSC) collaboration includes the astronomical communities of Japan and Taiwan, and Princeton University. The HSC instrumentation and software were developed by National Astronomical Observatory of Japan (NAOJ), Kavli Institute for the Physics and Mathematics of the Universe (Kavli IPMU), University of Tokyo, High Energy Accelerator Research Organization (KEK), Academia Sinica Institute for Astronomy and Astrophysics in Taiwan (ASIAA) and Princeton University.  
Funding was contributed by the FIRST program from Japanese Cabinet Office, Ministry of Education, Culture, Sports, Science and Technology (MEXT), Japan Society for the Promotion of Science (JSPS), Japan Science and Technology Agency (JST), Toray Science Foundation, NAOJ, Kavli IPMU, KEK, ASIAA and Princeton University.

% DECaLS part
The Legacy Surveys consist of three individual and complementary projects: the Dark Energy Camera Legacy Survey (DECaLS; Proposal ID \#2014B-0404; PIs: David Schlegel and Arjun Dey), the Beijing-Arizona Sky Survey (BASS; NOAO Prop. ID \#2015A-0801; PIs: Zhou Xu and Xiaohui Fan), and the Mayall z-band Legacy Survey (MzLS; Prop. ID \#2016A-0453; PI: Arjun Dey). DECaLS, BASS and MzLS together include data obtained, respectively, at the Blanco telescope, Cerro Tololo Inter-American Observatory, NSF’s NOIRLab; the Bok telescope, Steward Observatory, University of Arizona; and the Mayall telescope, Kitt Peak National Observatory, NOIRLab. The Legacy Surveys project is honored to be permitted to conduct astronomical research on Iolkam Du’ag (Kitt Peak), a mountain with particular significance to the Tohono O’odham Nation.

NOIRLab is operated by the Association of Universities for Research in Astronomy (AURA) under a cooperative agreement with the National Science Foundation.

This project used data obtained with the Dark Energy Camera (DECam), which was constructed by the Dark Energy Survey (DES) collaboration. Funding for the DES Projects has been provided by the U.S. Department of Energy, the U.S. National Science Foundation, the Ministry of Science and Education of Spain, the Science and Technology Facilities Council of the United Kingdom, the Higher Education Funding Council for England, the National Center for Supercomputing Applications at the University of Illinois at Urbana-Champaign, the Kavli Institute of Cosmological Physics at the University of Chicago, Center for Cosmology and Astro-Particle Physics at the Ohio State University, the Mitchell Institute for Fundamental Physics and Astronomy at Texas A\&M University, Financiadora de Estudos e Projetos, Fundacao Carlos Chagas Filho de Amparo, Financiadora de Estudos e Projetos, Fundacao Carlos Chagas Filho de Amparo a Pesquisa do Estado do Rio de Janeiro, Conselho Nacional de Desenvolvimento Cientifico e Tecnologico and the Ministerio da Ciencia, Tecnologia e Inovacao, the Deutsche Forschungsgemeinschaft and the Collaborating Institutions in the Dark Energy Survey. The Collaborating Institutions are Argonne National Laboratory, the University of California at Santa Cruz, the University of Cambridge, Centro de Investigaciones Energeticas, Medioambientales y Tecnologicas-Madrid, the University of Chicago, University College London, the DES-Brazil Consortium, the University of Edinburgh, the Eidgenossische Technische Hochschule (ETH) Zurich, Fermi National Accelerator Laboratory, the University of Illinois at Urbana-Champaign, the Institut de Ciencies de l’Espai (IEEC/CSIC), the Institut de Fisica d’Altes Energies, Lawrence Berkeley National Laboratory, the Ludwig Maximilians Universitat Munchen and the associated Excellence Cluster Universe, the University of Michigan, NSF’s NOIRLab, the University of Nottingham, the Ohio State University, the University of Pennsylvania, the University of Portsmouth, SLAC National Accelerator Laboratory, Stanford University, the University of Sussex, and Texas A\&M University.

BASS is a key project of the Telescope Access Program (TAP), which has been funded by the National Astronomical Observatories of China, the Chinese Academy of Sciences (the Strategic Priority Research Program ``The Emergence of Cosmological Structures'' Grant \# XDB09000000), and the Special Fund for Astronomy from the Ministry of Finance. The BASS is also supported by the External Cooperation Program of Chinese Academy of Sciences (Grant \# 114A11KYSB20160057), and Chinese National Natural Science Foundation (Grant \# 11433005).

The Legacy Survey team makes use of data products from the Near-Earth Object Wide-field Infrared Survey Explorer (NEOWISE), which is a project of the Jet Propulsion Laboratory/California Institute of Technology. NEOWISE is funded by the National Aeronautics and Space Administration.

The Legacy Surveys imaging of the DESI footprint is supported by the Director, Office of Science, Office of High Energy Physics of the U.S. Department of Energy under Contract No. DE-AC02-05CH1123, by the National Energy Research Scientific Computing Center, a DOE Office of Science User Facility under the same contract; and by the U.S. National Science Foundation, Division of Astronomical Sciences under Contract No. AST-0950945 to NOAO.

We gratefully acknowledge NSF grant AST-1613582, which supports the Dragonfly Wide Field Survey.

% SDSS part
Funding for the SDSS and SDSS-II has been provided by the Alfred P. Sloan Foundation, the Participating Institutions, National Science Foundation, the U.S. Department of Energy, National Aeronautics and Space Administration, the Japanese Monbukagakusho Scholarship, the Max Planck Society, and the Higher Education Funding Council for England. The SDSS is managed by the Astrophysical Research Consortium for the Participating Institutions. The Participating Institutions are the American Museum of Natural History, Astrophysical Institute Potsdam, University of Basel, University of Cambridge, Case Western Reserve University, University of Chicago, Drexel University, Fermilab, the Institute for Advanced Study, the Japan Participation Group, Johns Hopkins University, the Joint Institute for Nuclear Astrophysics, the Kavli Institute for Particle Astrophysics and Cosmology, the Korean Scientist Group, the Chinese Academy of Sciences (LAMOST), Los Alamos National Laboratory, the Max Planck Institute for Astronomy (MPIA), the Max Planck Institute for Astrophysics (MPA), New Mexico State University, Ohio State University, University of Pittsburgh, University of Portsmouth, Princeton University, the U.S. Naval Observatory, and University of Washington.  

% Packages
This research made use of: \href{http://www.numpy.org}{\code{NumPy}}, a fundamental package for scientific computing with Python \citep{numpy}; \href{https://www.scipy.org}{\code{SciPy}}, an open source scientific tool for Python \citep{scipy}; \href{https://matplotlib.org}{\code{Matplotlib}}, a 2-D plotting library for Python \citep{matplotlib}; \href{http://www.astropy.org}{\code{Astropy}}, a community-developed core Python package for Astronomy \citep{astropy2013}; \href{https://sep.readthedocs.io/en/v1.0.x/}{\code{sep}}, a Python library for Source Extraction and Photometry \citep{Bertin1996, Barbary2016}; \href{http://www.astromatic.net/software/swarp}{\code{SWarp}}, a program that resamples and co-adds FITS images using any arbitrary astrometric projection defined in the WCS standard \citep{Swarp}; \href{http://www.astromatic.net/software/scamp}{\code{SCAMP}}, a program for automatic astrometric and photometric calibration \citep{SCAMP}; \href{https://github.com/bd-j/sedpy}{\code{sedpy}}, a Python package for astronomical SEDs; \href{https://github.com/dstndstn/tractor}{\code{The Tractor}}, a tool for optimizing or sampling from models of astronomical objects \citep{Lang2016}; \href{http://galsim-developers.github.io/GalSim/_build/html/index.html}{\code{GalSim}}, the modular galaxy image simulation toolkit \citep{GalSim2015}; \href{https://photutils.readthedocs.io/en/stable/}{\code{Photutils}}, an Astropy package for detection and photometry of astronomical sources \citep{photutils}; \href{https://github.com/AstroJacobLi/mrf}{\code{mrf}}, a method for isolating faint, extended emission in Dragonfly data and other low-resolution images \citep{MRF}; \href{https://iraf-community.github.io/}{\code{IRAF}}, the Image Reduction and Analysis Facility \citep{IRAF1986,IRAF1993}. 

\section*{Data availability}
The data underlying this article will be shared on reasonable request to the corresponding author.

%\href{http://montage.ipac.caltech.edu}{\code{Montage}}, an astronomical image mosaic engine, which is funded by the National Science Foundation under Grant Number ACI-1440620, and was previously funded by the National Aeronautics and Space Administration's Earth Science Technology Office, Computation Technologies Project, under Cooperative Agreement Number NCC5-626 between NASA and the California Institute of Technology.

\appendix
\section{Mock Tests on HSC Sky Background}\label{appendix:mock-test}

The mock test results shown in Section \ref{sec:bkg-mock-test} justify our methodology for extracting surface brightness profiles from HSC images. In this appendix, we briefly summarize the {\skyobj} detection method, ways to use {\skyobj} to correct sky background residuals in HSC data release \code{S18A}, and caveats.

\begin{figure*}
	%\hskip -5mm
	\vbox{ 
		%\vskip -10mm
		\centering
		\includegraphics[width=1\linewidth]{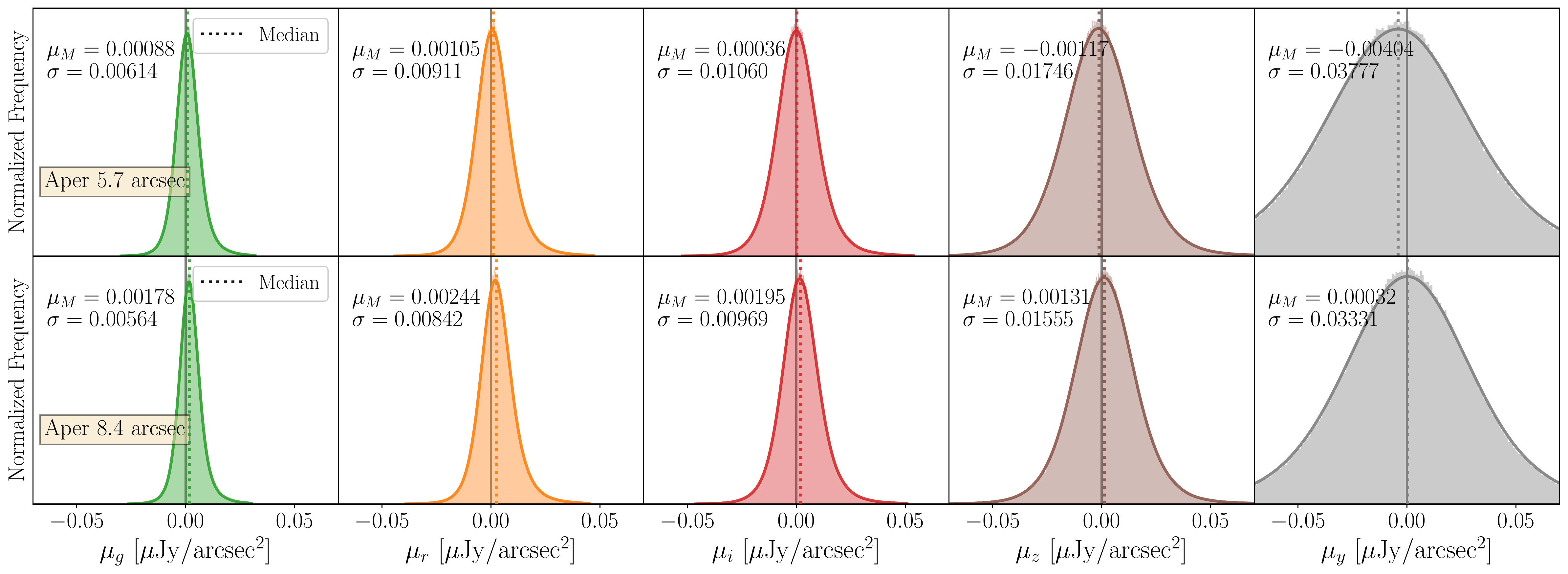}
	}
	\caption{Distribution of fluxes within \skyobjs{} in HSC \code{S18A} data. We show the 5.7 (8.4) arcsec \skyobjs{} in the upper (lower) panel. The distributions correspond to the $grizy$ filters from left to right. The median value ($\mu_M$) and the standard deviation ($\sigma$) of the distribution are shown. For 5.7 arcsec \skyobjs{}, $i$-band has the best global background subtraction, whereas the backgrounds in other bands are either over- or under-subtracted. For 8.4 arcsec \skyobjs{}, the backgrounds in all bands are under-subtracted. On the other hand, the scatter of \skyobjs{} flux increases from $g$-band to $y$-band. 
	}
	\label{fig:skyobj_stats}
\end{figure*}

\texttt{hscPipe} (version 6.7) detects void areas in single CCD images and uses different aperture sizes to evaluate the flux in these areas. \texttt{hscPipe} first masks out pixels with detections, then randomly selects void regions in the remaining sky; the regions are called {\skyobjs}. The pipeline picks 100 random {\skyobjs} in one patch (about $4000\times4000$ pixels, i.e. $12\arcmin\times12\arcmin$). There are six options for aperture diameters: $2.0\arcsec$, $3.0 \arcsec$, $4.0 \arcsec$, $5.7 \arcsec$, $8.4 \arcsec$ and $11.8 \arcsec$. The flux inside the sky object is a good proxy for the local sky background value. In Figure \ref{fig:skyobj_stats}, we show the distribution of fluxes of 5.7 arcsec and 8.4 arcsec \skyobjs{} in all HSC \code{S18A} data, as a function of filters. For the 5.7 arcsec \skyobjs{}, the $g$- and $r$-bands ($z$- and $y$-bands) on average show positive (negative) average global background values, while the $i$-band has the most accurate background subtraction. However, when using 8.4 arcsec aperture, all bands show positive global background residuals. Furthermore, the scatter increases monotonically with effective wavelength from $g$-band to $z$-band. 

When using {\skyobj} to characterize the sky around any given galaxy, there are two free parameters: aperture size and matching range. In the following tests, we show that matching $8.4 \arcsec$ {\skyobjs} located between $1\arcmin$ and $4 \arcmin$ from the target galaxy gives the best results for HSC \code{S18A} data. \skyobjs{} are matched using circular annulus. 

\begin{figure*}
	%\hskip -5mm
	\vbox{ 
		%\vskip -10mm
		\centering
		\includegraphics[width=0.94\linewidth]{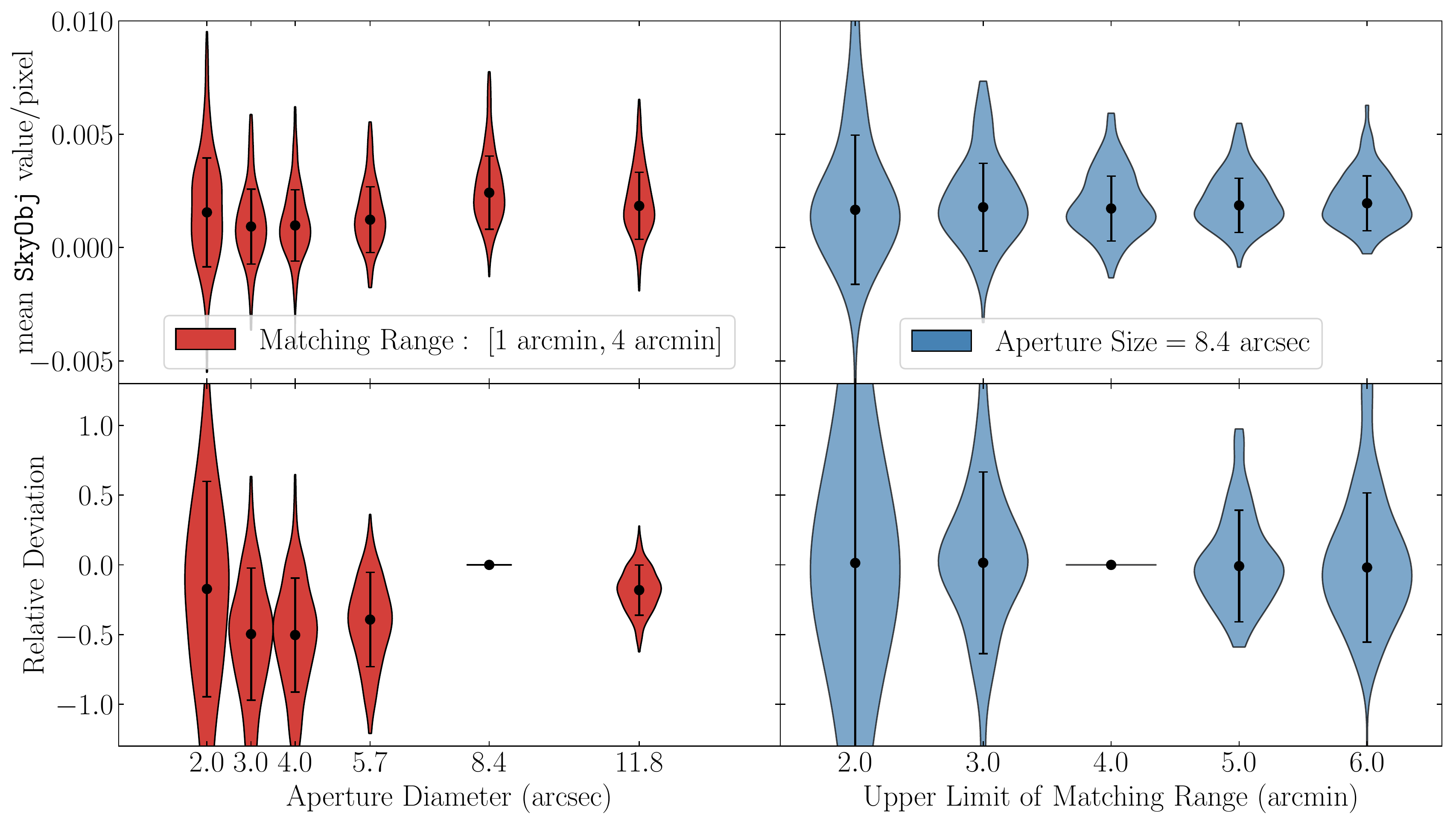}
	}
	\caption{{\skyobj} analysis results. We test the impact of aperture size and matching range. This figure shows that the mean value of {\skyobj} around an object does not depend much on matching range, but changes significantly with aperture size. Hence, we use the mock test results to test which aperture size is more reasonable, as shown in Figure \ref{fig:mock-test-aperture}.}
	\label{fig:skyobj-analysis}
\end{figure*}

We randomly select 300 {\skyobjs} positions to inject mock galaxies. To evaluate the local sky value around these fake galaxies, we match the surrounding {\skyobjs} and calculate their mean values. First, we study how the matching range affects the mean value of {\skyobjs}, with a fixed aperture size of $8.4\ \arcsec$. Since \skyobjs{} that are close to the galaxy might be contaminated by the light of the galaxy itself, we set the inner bound of matching range to $1 \arcmin$ radius. This excludes {\skyobjs} within 190 kpc for $z=0.19$ and 370 kpc for $z=0.50$, which is far enough from the central galaxy. We vary the upper bound of matching from $2 \arcmin$ to $6 \arcmin$. To exclude outlying {\skyobjs}, we execute a five-time sigma-clipping with a $3\sigma$ threshold. The blue violin plots in Figure \ref{fig:skyobj-analysis} show the distributions of mean {\skyobjs} values within different matching ranges. The medians stay nearly constant with the change of matching range, indicating that matching range does not affect the sky background correction too much. \skyobjs that are matched within [$1 \arcmin$, $2 \arcmin$] have large scatter, which might be caused by the small number of {\skyobjs} within this range. With the angular size of a single CCD of HSC about $6 \arcmin \times 12 \arcmin$, the larger matching range does not make sense because the variation of sky background is typically at the CCD-level. We fix the matching range to [$1\arcmin$, $4 \arcmin$] in the following tests.

We then test the effect of different aperture sizes. The red violin plots in Figure \ref{fig:skyobj-analysis} show the distributions of the mean {\skyobjs} value per pixel and their relative deviations with respect to the mean {\skyobjs} value of $8.4 \arcsec$ aperture size. The black dot shows the median of the violin plot and its error bar represents $1\sigma$ deviation. The trends in the two left panels are the same: the mean {\skyobjs} value of $8.4 \arcsec$ aperture size is the largest. The relative deviations of mean {\skyobjs} values with different aperture sizes are significant, as shown in the lower left panel of Figure \ref{fig:skyobj-analysis}. 

We further analyze the mock test results from Section \ref{sec:bkg-mock-test} as follows. We subtract the profiles for 60 mock images with mean values of {\skyobj} with different aperture sizes, then plot the median profiles in Figure \ref{fig:mock-test-compare}. The mean value of {\skyobjs} increases as the aperture size increases, since larger {\skyobjs} include faint sources and light from nearby bright objects. This trend is consistent with the violin plot. By comparing the median profiles with the model profiles, we find that sky correction using $8.4 \arcsec$ aperture could recover the model best for relatively low-$z$ (brighter) galaxies, whereas $5.7 \arcsec$ aperture is a better choice for sky correction around galaxies at higher redshift (fainter). 

However, we use a simplistic galaxy model in these mock tests that may not represent real massive galaxies perfectly. In future iterations, we will develop more realistic galaxy models to improve the robustness of these mock test. On the other hand, in real data, sky subtraction will be affected by the injected mock galaxy. Therefore, a mock test shows only one aspect of sky subtraction, and tests on real data are needed to affirm the soundness of using {\skyobjs} to correct sky residuals. Figure \ref{fig:mid-z-linear} in Section \ref{subsec:linear-profile} shows converged median profiles and symmetric distribution around zero intensity when the sky is corrected using an $8.4 \arcsec$ aperture {\skyobjs}. We also tested correcting sky residuals by using a $5.7\arcsec$ aperture {\skyobjs} on the intermediate-$z$ sample, but we found a slight under-subtraction. This proves that $8.4 \arcsec$ aperture size is the most reasonable choice for intermediate-$z$ galaxies regardless of stellar mass, since the median profiles converge to zero smoothly without any sign of over- or under-subtraction. 

Although \code{S18A} avoids over-subtraction around intermediate-$z$ galaxies, the sky residual makes the footprints of bright objects too large, which slows down the deblender dramatically, causes over-deblending issues and sometimes crashes the pipeline. If we use a large bright object mask that is too large, we see a drastic decrease of the galaxy sample size, which will impact cosmology studies using massive galaxies. In future data releases, \code{hscPipe} will detect objects on locally subtracted images and make catalogs, but the coadded globally subtracted images will still be retained for future use. 

\begin{figure*}
	%\hskip -5mm
	\vbox{ 
		%\vskip -10mm
		\centering
		\includegraphics[width=1\linewidth]{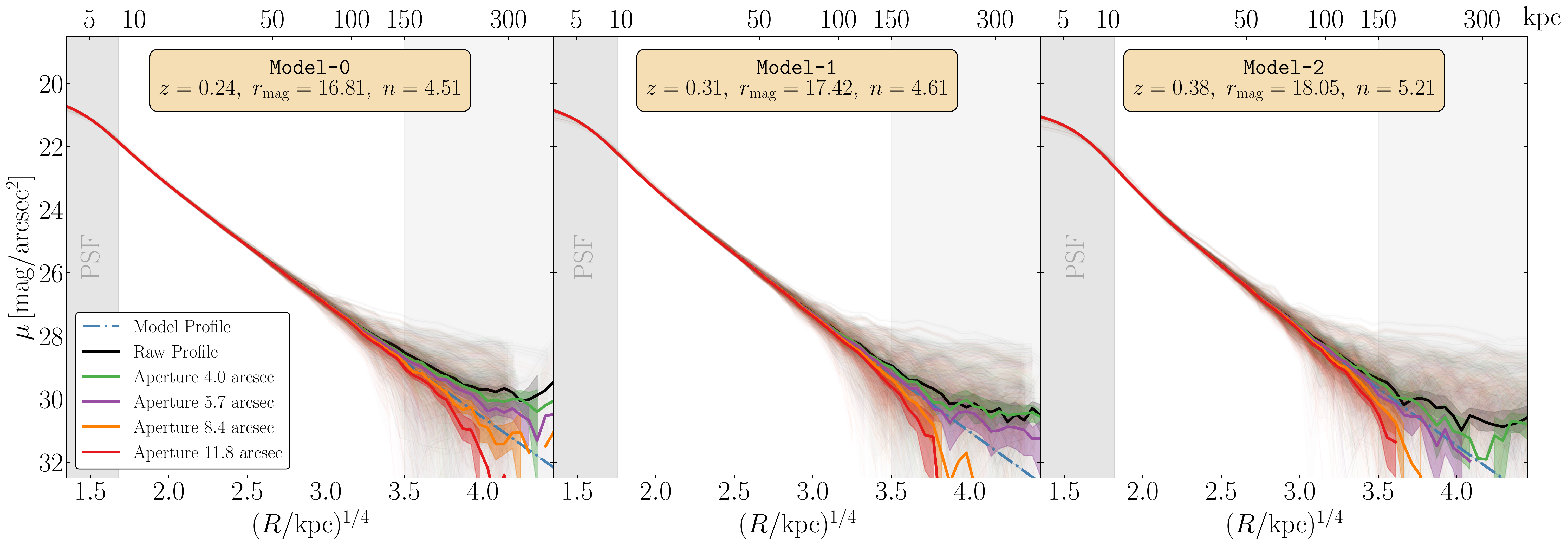}
	}
	\caption{Median profiles of mock test profiles after we subtract the mean value of {\skyobj} with different aperture size. The matching range is fixed to $1 \arcmin < D < 4 \arcmin$. This figure clearly shows the behaviors and differences between methods and suggests that subtracting the mean value of {\skyobj} with $8.4 \arcsec$ aperture (for relatively low redshift) and $5.7 \arcsec$ aperture (for relatively high redshift) give us the best results.}
	\label{fig:mock-test-aperture}
\end{figure*}

\section{Modeling Neighboring objects Instead of Masking}\label{subsec:2Dmodel}
Satellite galaxies, stars, and foreground/background galaxies contaminate the surface brightness profiles of target galaxies. In Section \ref{sec:SBP-method}, we describe how we use a binary mask to remove these contaminants. We also perform sigma clipping along the isophote and take the median intensity value to reduce the impact of unmasked fluxes of contaminants. We have performed tests using synthetic galaxies to show that we can recover profiles accurately using this binary mask. Meanwhile, another method is to model surrounding objects using 2-D models and to subtract these models from the image, leaving the target galaxy in the center. This method is used in DECaLS data releases as well as the customized pipeline used in this work (Section \ref{sec:decals_methods}). Here in this appendix, we test to which extent the 2-D modeling method agree with the binary mask method using mock galaxies injected in HSC images.

\begin{figure*}
	%\hskip -5mm
	\vbox{ 
		%\vskip -10mm
		\centering
		\includegraphics[width=1\linewidth]{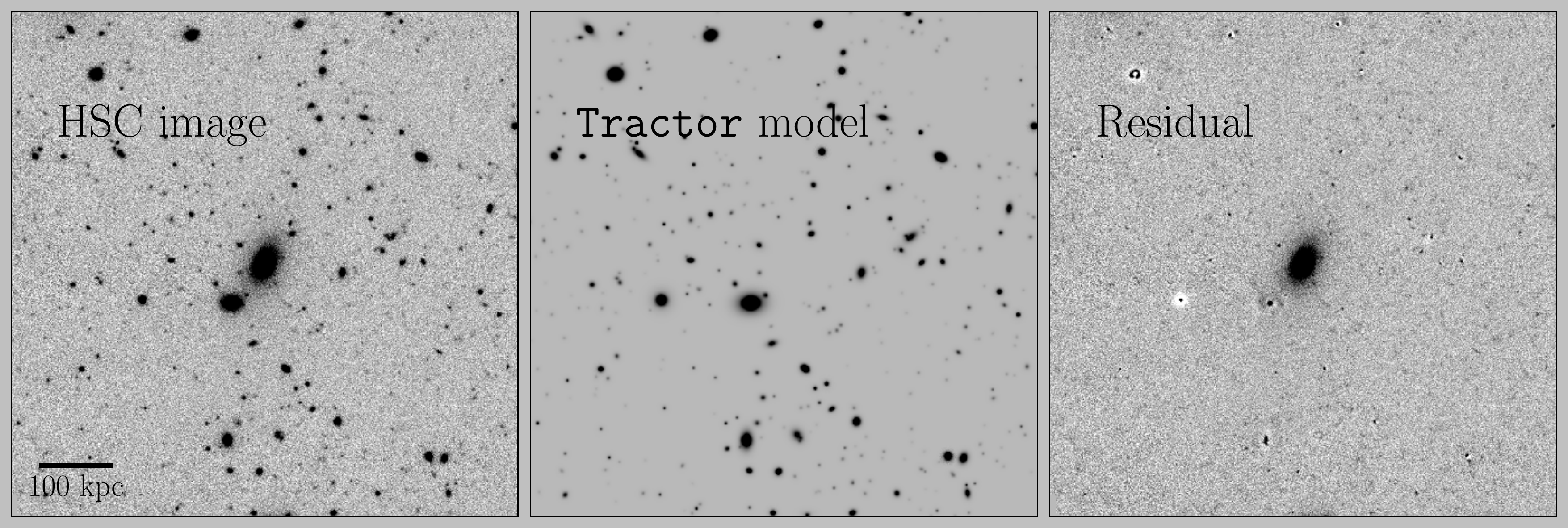}
	}
	\caption{Modeling and subtracting neighboring stars and galaxies with {\tractor}. The central galaxy is drawn from the intermediate-$z$ sample and has $z=0.294,\ r_{\mathrm{cModel}}=18.17\ \mathrm{mag},\ \log M_{*}^{\mathrm{max}}/M_\odot = 11.67$. The middle panel shows {\tractor} models for surrounding galaxies (excluding the central galaxy). The right panel shows the image after subtracting neighbors. Small residuals are masked out before extracting 1-D surface brightness profiles.}
	\label{fig:tractor_demo}
\end{figure*}

\code{The Tractor}\footnote{\url{https://github.com/dstndstn/tractor}} is a probability-based photometry and model fitting tool \citep{Lang2016}, which has been successfully used in DECaLS data releases. Here we use {\tractor} to build 2-D models of neighboring objects and subtract them from HSC images. We first use an HSC bright star mask to mask out bright or saturated stars that cannot be well-described by simple models. Then all the objects in the image are extracted by {\sep}, along with their shapes and fluxes. We need to assign prior model types to objects to run {\tractor} (point source, de Vaucouleurs profile, exponential profile and composite profile). DECaLS provides {\tractor} fitting results (e.g., model type, shape, flux) for each object within its footprint. We match objects in the image with the DECaLS catalogs and use the model types drawn from DECaLS. The shapes and fluxes of the models are, however, still based on our measurements using {\sep}. For objects that are not matched in DECaLS catalogs, we assign their type to be `de Vaucouleurs'. The central target object is assigned to be `composite'. Fitting hundreds of objects together at the same time is very time consuming. For this reason, we first fit the 40 brightest objects together, then add more objects to the optimized model and repeat the fitting. This process is repeated until the model of every object is added and optimized. Figure \ref{fig:tractor_demo} demonstrates an example of modeling using {\tractor}. The satellites, stars, and other galaxies around the target galaxy have been modeled and removed in the right panel. Nevertheless, many galaxies cannot be simply modeled by `de Vaucouleurs' or `exponential', as residuals show up as bright dots and negative rings. We mask out the remaining residuals before extracting surface brightness profiles of the central target.

\begin{figure}
	\centering
	\includegraphics[width=1\linewidth]{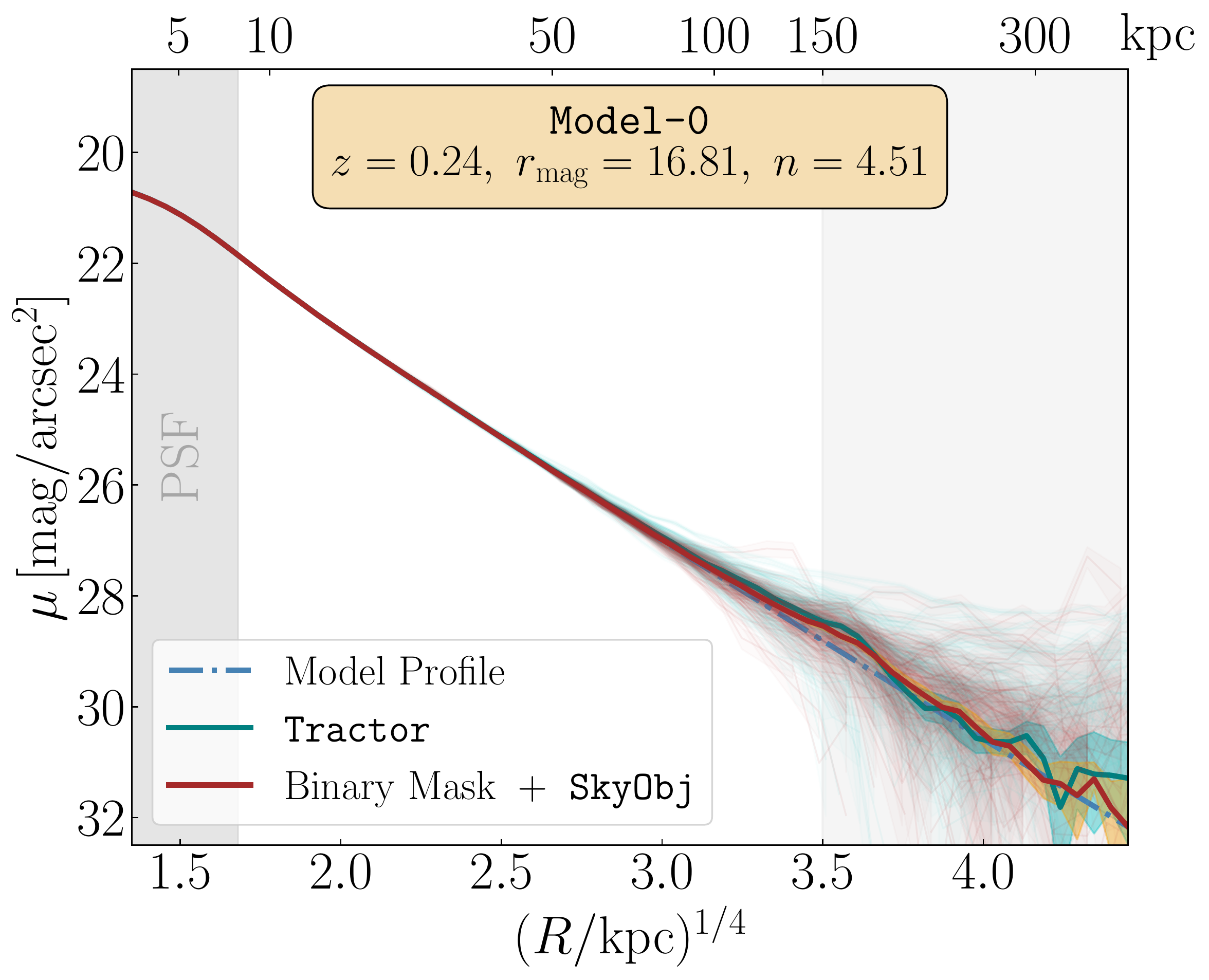}
	\vspace{-20pt}
	\caption{Median profile of mock galaxies after subtracting neighboring stars and galaxies using {\tractor} (green line). The red line shows the results of the binary mask with the {\skyobj} correction. The two methods yield the same results but {\tractor} takes much more time to run.}
	\label{fig:tractor_mock_test}
\end{figure}

Although {\tractor} behaves well for individual galaxies, it is not clear that it will improve the measurement of surface brightness profiles -- mock tests are required. Following the same procedure as in Section \ref{sec:bkg-mock-test}, we generate 70 mock images with the first model ($z=0.24$, $r_{\mathrm{mag}} = 16.81$) in $r$-band. Then we subtract neighbors using {\tractor} and extract surface brightness profiles. The 2-D modeling method also subtracts the residual sky background from the image. Figure \ref{fig:tractor_mock_test} shows that 2-D modeling with the {\tractor} and the binary mask method yields consistent results. However, the 2-D method is more computationally expensive and so we adopt the binary mask method for the remainder of this paper.

In a companion paper \citep[][]{Ardila2021}, we have also tested satellite removal using mock images from Illustris \citep{Vogelsberger+2014:Illustris, Vogelsberger+2014:Nature} and IllustrisTNG \citep{Springel+2018:TNG, Pillepich+2018b:TNG} simulations. The results of those tests show that residual light from satellites has a negligible impact on 1-D surface brightness profiles. 

\section{Filter Corrections}\label{subsec:filt-corr}
To compare the photometry of HSC, DECaLS, SDSS and Dragonfly, we need to account for filter differences. In this appendix, we use synthetic stellar photometry to derive a filter correction scheme for the $r$-bands. Given the response curves of each band and a spectrum, we calculate the AB magnitude in each band for stars. We use spectra from the Gunn-Stryker stellar spectra atlas \citep{Gunn-Stryker1983} and calculate AB magnitudes using \href{https://github.com/bd-j/sedpy}{\code{sedpy}}. By assuming that galaxies and stars with the same color have \textit{nearly} the same spectral energy distribution (SED), we obtain the filter correction for each galaxy. 

Taking the intermediate-$z$ sample as an example, we fit a third-order polynomial between the source color in HSC $g_{\mathrm{HSC}} - i_{\mathrm{HSC}}$ and magnitude difference $\Delta_r = r_{\mathrm{HSC}} - r_{\mathrm{DECaLS}}$ between HSC and DECaLS. The best-fit polynomial is 
\begin{equation}
    \Delta_r(x) = 0.0024 x^3 - 0.0097 x^2 + 0.0707 x + 0.0004,
\end{equation}
where $\Delta_r = r_{\mathrm{HSC}} - r_{\mathrm{DECaLS}}$ and $x = g_{\mathrm{HSC}} - i_{\mathrm{HSC}}$. Then we use the HSC color ($g_{\mathrm{HSC}} - i_{\mathrm{HSC}}$, using \code{cModel} magnitudes) to calculate the magnitude difference $\Delta_r$ according to the fitted polynomial and add it to the DECaLS profile. For the low-$z$ sample, we calibrate other surveys to SDSS in the same way, but using $g_{\mathrm{SDSS}} - r_{\mathrm{SDSS}}$ colors instead. The filter correction term is typically around 0.15 mag for the intermediate-$z$ sample and is less than 0.10 mag for the low-$z$ sample. We have applied filter corrections to all profiles presented in this paper. 

This method, however, has limitations. It assumes that the SED of a galaxy is the same as the SED of a star with the same color, which is not necessarily true. A more detailed analysis using SED fitting is needed to derive the filter correction term for each galaxy. Furthermore, we ignore the color gradient in a galaxy when applying the filter correction term to the surface brightness profile. These corrections are, however, beyond the scope of this paper.

 %-------------- BIBLIOGRAPHY -------------------------------------------------------

\bibliographystyle{mnras}
\bibliography{citation}

 %-----------------------------------------------------------------------------
\label{lastpage}
\end{document}

%% file: define.tex
\newcommand{\code}[1]{\textbf{\texttt{#1}}}
\newcommand{\tractor}{\code{the Tractor}}
\newcommand{\sep}{\code{sep}}
\newcommand{\ellipse}{\code{Ellipse}}
\newcommand{\skyobj}{\code{SkyObj}}
\newcommand{\skyobjs}{\code{SkyObjs}}
\newcommand{\galsim}{\code{GalSim}}
\newcommand{\sersic}{S\'ersic}